\documentclass[nofootinbib,prd,superscriptaddress,showpacs,preprint]{revtex4}
\usepackage{chay,epsfig}
 
\newcommand{\w}{\overline{\omega}} 

\begin{document}
\title{Deep inelastic scattering near the endpoint \\ in 
  soft-collinear effective theory}
\author{Junegone Chay}\email{chay@korea.ac.kr} 
\affiliation{Department of Physics, Korea University, Seoul 136-701,
Korea}

\author{Chul Kim}\email{chk30@pitt.edu}
\affiliation{Department of Physics and Astronomy, University of
  Pittsburgh, PA 15260, U.S.A.} 

\preprint{}
\begin{abstract}
We apply the soft-collinear effective theory (SCET) to deep inelastic
scattering near the endpoint region. The forward scattering amplitude,
and the structure functions are shown to factorize as a 
convolution of the Wilson coefficients, the jet functions, the parton
distribution functions. The behavior of the parton distribution
functions near the endpoint region is considered. It turns out that it
evolves with the Altarelli-Parisi kernel even in the endpoint region,
and the parton distribution function can be factorized further into a
collinear part and the soft Wilson line. 
The factorized form for the structure functions is obtained by the
two-step matching, and the radiative corrections or the 
evolution for each factorized part can be computed in perturbation
theory. We present the radiative corrections of each factorized part
to leading order in $\alpha_s$, including the zero-bin subtraction for
the collinear part.   
\end{abstract}

\maketitle 

\section{Introduction}
The soft-collinear effective theory
(SCET)\cite{Bauer:2000ew,Bauer:2000yr,Bauer:2001yt}  is a useful
theoretical tool to treat physical processes with energetic light
particles in a systematic way. For an energetic particle moving in the
$n^{\mu}$ direction, the momentum can be decomposed into
\begin{equation}
  \label{como}
  p^{\mu} =\frac{\overline{n}\cdot p}{2} n^{\mu} +p_{\perp}^{\mu}
  +\frac{n\cdot p}{2} \overline{n}^{\mu} \sim \mathcal{O} (Q)
  +\mathcal{O} (\Lambda) +\mathcal{O} (\Lambda^2/Q),
\end{equation}
where $Q$ is a large scale, and $n^{\mu}$, $\overline{n}^{\mu}$ are
lightlike vectors satisfying 
$n^2 =\overline{n}^2 =0$, $n\cdot \overline{n} =2$. Each component has
a distinct scale in powers of $\Lambda$ which is a typical hadronic
scale, and SCET describes the interactions of the collinear
particles and the ultrasoft (usoft) particles with momentum
$p_{us}^{\mu} \sim (\Lambda, \Lambda,\Lambda)$. 
Since there are three distinct scales for the momentum
of a collinear particle, SCET employs a two-step
matching process by integrating out large energy scales
successively \cite{Bauer:2001yt}. In the first stage the degrees of
freedom of order $Q$ from the full theory are integrated out to
produce $\mathrm{SCET}_{\mathrm{I}}$. In $\mathrm{SCET}_{\mathrm{I}}$,  
collinear particles are allowed to interact with usoft particles and
the typical virtuality of the collinear particles is
$p^2_{hc} \sim Q\Lambda$. In the second stage the degrees of freedom
with $p^2 \sim Q\Lambda$ are integrated out, and the remaining
effective theory in which all the particles have $p^2 \sim \Lambda^2$
is called $\mathrm{SCET}_{\mathrm{II}}$. Here the collinear particles
are decoupled from the soft particles. The Wilson coefficients of
operators and the renormalization behavior of them can be 
computed perturbatively by matching the effective theories at each
boundary. 

SCET has been successfully applied to various $B$ meson decays
\cite{Chay:2002vy, Htl, Leptonic, Chay:2003ju, Nonleptonic,
  Mantry:2003uz,Chay:2003kb,Bauer:2000ew,Inclusive}. It
is especially convenient to study the factorization properties of
$B$ decays including spectator interactions since SCET is formulated
such that soft and collinear particles are decoupled. On the other
hand, SCET can be applied to other high-energy processes which include 
energetic light particles \cite{Bauer:2002nz,
Quarkonium, Chay:2004zn, Manohar:2003vb,Jet}. 
It has been applied  to deep inelastic scattering (DIS) near the
endpoint region using SCET \cite{Manohar:2003vb,Becher:2006mr} and
using an effective theory scheme \cite{Chen:2006vd}. 

In this paper we analyze the endpoint region in DIS more carefully
using the two-step matching to show the explicit factorization of the
structure functions in terms of the hard part, the jet function, the
soft gluon emissions, and the collinear matrix elements. We also
discuss and compare delicate physical meanings and implications of the
parton distribution functions in the endpoint region, defined both in
the full theory and in SCET. 
In addition to showing the factorization, we take one step further to
consider another aspect of DIS, namely the behavior of the
longitudinal structure function near the endpoint region. The
longitudinal structure function 
vanishes at leading order in $\alpha_s$ due to the fact that the
parton (quark) in the proton has spin 1/2. However this is broken at
order $\alpha_s$ and the longitudinal structure function is further
suppressed by $\Lambda/Q$, which we explicitly present here.   

In Section \ref{kinematics} we explain the kinematics of DIS. We
choose the Breit frame and present how the momenta scale in powers
of $\Lambda$, which is useful in constructing and matching effective
theories. The forward scattering amplitude, and the structure
functions are defined in SCET, compared with those in the full
theory. In Section \ref{opscet} the method to compute the forward
scattering amplitudes in DIS using SCET is described. The
leading and the subleading currents are introduced and the
prescription for the  usoft factorization is explained. In Section
\ref{f1}, we compute the structure function $F_1 (x,Q)$, and show that
it factorizes. In Section \ref{pdfs}, we consider the parton
distribution near the endpoint region, and express the forward
scattering amplitude in terms of the parton distribution function. In
Section \ref{moana}, we present the moments of the structure functions
as a product of the moments for each factorized term. In Section
\ref{radcor}, we compute the radiative corrections of each factorized
term to order in $\alpha_s$, and express the moments of the structure
functions to leading logarithmic accuracy. In Section \ref{fl}, we
compute the longitudinal structure function $F_L (x,Q)$ in SCET and
show that it also factorizes. In the final section, we give a
conclusion. In Appendix A, the zero-bin subtraction method
\cite{Manohar:2006nz} in $\mathrm{SCET}_{\mathrm{I}}$ before the usoft
factorization of the collinear fields is discussed. In Appendix B, the
procedure for taking the imaginary part in
$\mathrm{SCET}_{\mathrm{I}}$ and $\mathrm{SCET}_{\mathrm{II}}$ is
explained. In Appendix C, the anomalous dimension of the operator
$J_{\mu}^{(1b)}$ is computed to order $\alpha_s$.  

\section{Kinematics\label{kinematics}} 
Let us consider the electroproduction in DIS $ep
\rightarrow e X$ near the endpoint region. The hadronic process
consists of $\gamma^* p \rightarrow X$, and we choose the Breit frame
in which the incoming proton is in the $\overline{n}^{\mu}$ direction,
and the outgoing hadrons are mainly in the $n^{\mu}$ direction. The
momentum transfer $q^{\mu}$ from the leptonic system is given by
\begin{equation}
q^{\mu} = (\overline{n} \cdot q, q_{\perp}^{\mu}, n\cdot q) =
(Q,0,-Q) = \frac{Q}{2} (n^{\mu}-\overline{n}^{\mu}), 
\end{equation}
where $q^2 = -Q^2$ is the large scale. The Bjorken variable $x$ is
defined as 
\begin{equation}
  x= \frac{Q^2}{2P\cdot q} \sim \frac{Q}{n\cdot P}, 
\end{equation}
where $P^{\mu}$ is the proton momentum in the $\overline{n}^{\mu}$
direction. The momenta of the proton $P^{\mu}$ and the final-state
particles $p_X=P+q$ are given by 
\begin{eqnarray}
P^{\mu} &=& (\overline{n}\cdot P, P_{\perp}^{\mu}, n\cdot P) \sim
\Bigl( \frac{x\Lambda^2}{Q}, P_{\perp}^{\mu}, \frac{Q}{x} \Bigr),
\nonumber  \\    
p_X^{\mu} &=&  (\overline{n} \cdot p_X, p_{X\perp}^{\mu}, n\cdot p_X)
\sim \Bigl(Q, p_{X\perp}^{\mu}, \frac{1-x}{x} Q \Bigr), 
\end{eqnarray}
with $P^2 \sim \Lambda^2$, $p_X^2 = Q^2 (1-x)/x$ where $\Lambda$ is a
typical hadronic scale of order 1 GeV. Near the endpoint where $x$
approaches 1 ($1-x \sim \Lambda/Q$)\footnote{In fact, $1-x$ does not
  have to be of order $\Lambda/Q$. Instead, we can introduce a small
  parameter $\delta = 1-x$ and $p_X^2 \sim Q \delta \gg
  \Lambda^2$. But for simplicity, we consider the case with $1-x \sim
  \Lambda/Q$.}, the invariant mass squared of the 
final-state particles becomes $p_X^2 \sim Q^2 (1-x) \sim
Q\Lambda$. Then the final-state particles can be regarded as 
collinear particles in $\mathrm{SCET}_{\mathrm{I}}$, which are
integrated out to obtain $\mathrm{SCET}_{\mathrm{II}}$ through the
two-step matching procedure. 

At the parton level, let $p^{\mu}$ be the momentum of the incoming
parton inside the proton, and let $y$ be the longitudinal momentum
fraction ($n\cdot p = y n\cdot P$). Then the partonic Bjorken variable
$w$ is given as
\begin{equation}
  w=\frac{Q^2}{2p\cdot q} \sim -\frac{n\cdot q}{n\cdot p}
  =\frac{x}{y}. 
\end{equation}
The momentum of the outgoing parton as $p^{\prime\mu}$ can be written
as 
\begin{equation}
p^{\prime\mu} = p^{\mu} + q^{\mu} = (\overline{n}\cdot p^{\prime},
p_{\perp}^{\prime\mu}, n\cdot p^{\prime}) \sim (Q,
p^{\prime\mu}_{\perp}, (1-w) n\cdot p).   
\end{equation}
And the endpoint region corresponds to $1-w \sim \Lambda/Q$ such that
$p^{\prime 2} \sim Q\Lambda$. 

The spin-averaged cross section for DIS can be
written as
\begin{equation}
  d\sigma = \frac{d^3 {\bf k}^{\prime}}{2|{\bf k}^{\prime}| (2\pi)^3}
  \frac{\pi e^4}{sQ^4} L^{\mu\nu} (k,k^{\prime}) W_{\mu\nu} (p,q), 
\end{equation}
where $k$ and $k^{\prime}$ are the incoming and outgoing lepton
momenta with $q=k^{\prime} -k$, $L^{\mu\nu}$ is the lepton tensor, and
$s=(p+k)^2$. The hadronic tensor $W_{\mu\nu}$ is related to the
imaginary part of the forward scattering amplitude $T^{\mu\nu}$.
The forward scattering amplitude is the spin-averaged
matrix element of the time-ordered product of the electromagnetic
currents, written as
\begin{equation}
T_{\mu\nu} (x,Q) = \langle P |\hat{T}_{\mu\nu}
|P\rangle_{\mathrm{spin\     av.}}, \ \ 
\hat{T}_{\mu\nu} (x,Q) =i\int d^4 z e^{iq\cdot z}
T\Bigl[J_{\mu}^{\dagger} (z) J_{\nu} (0)\Bigr],  
\end{equation}
where $J_{\mu}$ is the electromagnetic current. The relation between
the hadronic tensor $W_{\mu\nu}$ and the forward scattering amplitude
$T_{\mu\nu}$ is given by 
\begin{equation}
  W_{\mu\nu} (x,Q) =\frac{1}{\pi} \mathrm{Im} \, T_{\mu\nu} (x,Q).
\end{equation}

In electroproduction, considering all the possible Lorentz
structure,  $T_{\mu\nu}$ can be generally written as  
\begin{equation}
T_{\mu\nu} (x,Q) = -g^{\perp}_{\mu\nu} T_1  +
(\overline{n}_{\mu} n_{\nu} + \overline{n}_{\nu} n_{\mu}) T_2 +  
(\overline{n}_{\mu} n_{\nu} - \overline{n}_{\nu} n_{\mu}) T_3 +
\overline{n}_{\mu}\overline{n}_{\nu} T_4 +   n_{\mu}n_{\nu} T_5,
\end{equation}
where $g^{\perp}_{\mu\nu} = g_{\mu\nu} -(n_{\mu} \overline{n}_{\nu} +
\overline{n}_{\mu} n_{\nu})/2$. 
Due to the current conservation ($q_{\mu}T^{\mu\nu}=0)$, and the parity
conservation, we have $T_4=T_5=T_2$, and
$T_3=0$. Therefore the forward scattering amplitude has two
independent quantities, and is given by
\begin{equation} \label{tmunu1}
  T_{\mu\nu} (x,Q) = -g^{\perp}_{\mu\nu} T_1 (x,Q) +(n_{\mu}
  +\overline{n}_{\mu})(n_{\nu} +\overline{n}_{\nu}) T_2 (x,Q). 
\end{equation}
This can be cast into different forms using the fact that and any
terms proportional to $q_{\mu} =Q(n_{\mu} -\overline{n}_{\mu})/2$ can
be discarded since they vanish when they are contracted with the
lepton tensor. We can write $n_{\mu} +\overline{n}_{\mu} =
(\overline{n}_{\mu} -n_{\mu}) +2 n_{\mu}=(n_{\mu}
-\overline{n}_{\mu}) +2\overline{n}_{\mu}$ and drop the terms
proportional to $n_{\mu} 
-\overline{n}_{\mu}$. Then Eq.~(\ref{tmunu1}) can be equivalently
written as 
\begin{equation} \label{tmunu2}
  T_{\mu\nu} (x,Q) = -g^{\perp}_{\mu\nu} T_1 (x,Q) +4 n_{\mu} n_{\nu}
  T_2 (x,Q)  = -g^{\perp}_{\mu\nu} T_1 (x,Q) +
4 \overline{n}_{\mu} \overline{n}_{\nu} T_2 (x,Q). 
\end{equation}

The structure functions are defined from the hadronic tensor
$W_{\mu\nu}$ as 
\begin{equation} \label{conw}
  W_{\mu\nu} (x,Q) = -g_{\mu\nu} F_1 (x,Q) +\frac{P_{\mu}
  P_{\nu}}{P\cdot q} F_2 (x,Q),  
\end{equation}
where the terms proportional to $q_{\mu}$ or $q_{\nu}$ are dropped.
Using $P_{\mu} = n\cdot P \overline{n}_{\mu}/2$, $2P\cdot q =
n\cdot P \overline{n}\cdot q = Q^2 /x$, we can write
Eq.~(\ref{conw}) as 
\begin{eqnarray} \label{wmunus}
W_{\mu\nu} (x,Q) &=& -g_{\mu\nu} F_1 (x,Q) + \frac{\overline{n}_{\mu}
  \overline{n}_{\nu}}{2x} F_2 (x,Q) \nonumber \\
&=& -g^{\perp}_{\mu\nu} F_1 (x,Q) -\frac{1}{2}
(n_{\mu}\overline{n}_{\nu} +\overline{n}_{\mu} n_{\nu}) F_1 (x,Q)
+\frac{\overline{n}_{\mu}   \overline{n}_{\nu}}{2x} F_2 (x,Q)
\nonumber \\
&\longrightarrow& -g^{\perp}_{\mu\nu} F_1 (x,Q)
+\frac{\overline{n}_{\mu}   \overline{n}_{\nu}}{2} \Bigl( \frac{1}{x}
F_2 (x,Q) -2 F_1 (x,Q) \Bigr) \nonumber \\
&=&  -g^{\perp}_{\mu\nu} F_1 (x,Q) +\frac{\overline{n}_{\mu}
  \overline{n}_{\nu}}{2} F_L (x,Q),
\end{eqnarray}
where we extract $n_{\mu}-\overline{n}_{\mu}$ and discard it to obtain
the third relation, and the longitudinal structure function $F_L
(x,Q)$ is defined as
\begin{equation}
  F_L (x,Q) = \frac{1}{x}F_2 (x,Q) -2 F_1 (x,Q).
\end{equation}
The Lorentz structure $\overline{n}_{\mu} \overline{n}_{\nu}$ in
the final expression of Eq.~(\ref{wmunus}) can be replaced by
$n_{\mu}n_{\nu}$. Comparing Eqs.~(\ref{tmunu1}) and (\ref{wmunus}), we
obtain the relations
\begin{equation} \label{stt}
 F_1 (x,Q) = \frac{1}{\pi} \mathrm{Im} T_1 (x,Q), \ F_L (x,Q) =
 \frac{8}{\pi} \mathrm{Im} T_2 (x,Q).  
\end{equation}

As we will show explicitly, $T_1 (x,Q)$
receives the contribution at leading order, and $T_2 (x,Q)$ is
suppressed by $\Lambda/Q$ and $\alpha_s$ compared to $T_1
(x,Q)$. Therefore the Callan-Gross relation $F_L =0$ holds to leading
order, but is violated at subleading order. Here we also present $F_L$
computed using SCET. In fact, the equivalence between
Eq.~(\ref{tmunu1}) and Eq.~(\ref{tmunu2}) turns out to imply nontrivial
relations because the subleading contributions proportional to
$n^{\mu}n^{\nu}$ and $\overline{n}^{\mu} \overline{n}^{\mu}$ come from
different subleading current operators in SCET. The statement that
all the expressions are equivalent means 
that the longitudinal structure functions can be obtained using any
subleading current operators and it holds to all orders in
$\alpha_s$. The nontrivial relation will be verified in
this paper at order $\alpha_s$.  

\section{Operators  in SCET near the endpoint region\label{opscet}} 
In computing the forward scattering amplitude, we first express
the electromagnetic current $J_{\mu}$ in terms of the
effective fields in $\mathrm{SCET}_{\mathrm{I}}$. The
electromagnetic current operator at leading order in SCET is given by
\begin{equation} 
\overline{q} \gamma_{\mu} q \rightarrow C(Q,\mu) (J_{\mu}^{(0)}
+J_{\mu}^{(0)\dagger})
=C(Q, \mu) \Bigl[\overline{\xi}_{n}W_n \gamma_{\mu}^{\perp} 
W_{\overline{n}}^{\dagger} ~\xi_{\overline{n}}
+\overline{\xi}_{\bar{n}} W_{\bar{n}} \gamma_{\mu}^{\perp}
W_n^{\dagger} \xi_n\Bigr], 
\label{current} 
\end{equation}
where $\xi_n$ ($\xi_{\bar{n}}$) is the $n$ ($\overline{n}$) collinear
fermion field in SCET. Here $W_n$ and $W_{\bar{n}}$ are the collinear
Wilson lines 
\begin{equation} 
W_n (x)=\Bigl[\sum_{\rm{perms}} \exp \Bigl(-g 
\frac{1}{\overline{n}\cdot \mathcal{P}}
\overline{n} \cdot A_n (x) \Bigr) \Bigr], \
W_{\bar{n}} (x)=\Bigl[\sum_{\rm{perms}} \exp \Bigl(-g 
\frac{1}{n\cdot \mathcal{P}}
n \cdot A_{\bar{n}} (x) \Bigr) \Bigr].
\end{equation} 
Here $A_n^{\mu}$ ($A_{\bar{n}}^{\mu}$) is the collinear gluon in the
$n^{\mu}$ ($\overline{n}^{\mu}$) direction and the summation over the
label momenta is suppressed. The Wilson coefficient $C(Q,\mu)$ is
actually an operator and Eq.~(\ref{current}) is written as
\begin{eqnarray} \label{opcu}
&&\overline{\xi}_{n}W_n \gamma_{\mu}^{\perp} C(\overline{n}\cdot
\mathcal{P}^{\dagger}, n\cdot \mathcal{P})
W_{\overline{n}}^{\dagger} ~\xi_{\overline{n}} + 
\mathrm{h. c.} \nonumber \\
&&=  \int d\omega d\bar{\omega} C(\omega, \bar{\omega})
\overline{\xi}_{n}W_n \delta (\bar{\omega} -\overline{n} \cdot
\mathcal{P}^{\dagger}) \gamma_{\mu}^{\perp} \delta (\omega 
- n\cdot \mathcal{P}) W_{\overline{n}}^{\dagger} 
~\xi_{\overline{n}} +   \mathrm{h. c.},  
\end{eqnarray}
where $n\cdot \mathcal{P}$ ($\overline{n} \cdot
\mathcal{P}^{\dagger}$) is the operator extracting the label momentum in  
the $\overline{n}$ ($n$) direction. The operator form in
Eq.~(\ref{opcu}) is useful in deriving the Feynman rules to compute
radiative corrections. The hard coefficient $C(Q,\mu)$ can be
obtained from matching the full theory onto
$\mathrm{SCET}_{\mathrm{I}}$, it is given to order $\alpha_s$ as
\cite{Manohar:2003vb} 
\begin{equation} 
C(Q, \mu) = 1 + \frac{\alpha_s C_F}{4\pi} 
\Bigl( -\ln^2 \frac{Q^2}{\mu^2} + 3 \ln \frac{Q^2}{\mu^2} - 8 +
\frac{\pi^2}{6} \Bigr).
\label{Wilson} 
\end{equation} 
The hard coefficient  $C(Q,\mu)$ satisfies the renormalization group
equation 
\begin{equation} \label{evc}
  \mu\frac{d C(Q,\mu)}{d\mu} = \gamma_H (\mu) C(Q,\mu), \ \ 
  \gamma_H (\mu) = \frac{\alpha_s (\mu)C_F}{2\pi} \Bigl( 4\ln
  \frac{\mu}{Q} +3\Bigr).  
\end{equation}
 
We can obtain subleading current operators at order
$\sqrt{\Lambda/Q}$, which contain either $iD_n^{\perp}$ or
$iD_{\bar{n}}^{\perp}$. There are two independent operators involving
$iD_n^{\perp}$, one of which arises from the subleading correction
to the fermion field $\xi_{\bar{n}}$   
 \begin{equation}\label{expxi}
  \overline{q} \gamma_{\mu} q \rightarrow \overline{\xi}_n W_n \Bigl(1 
  +\frac{\FMslash{\overline{n}}}{2} W_n^{\dagger}
  i\overleftarrow{\FMSlash{D}_n^{\perp}} W_n \frac{1}{\overline{n}
    \cdot \mathcal{P}^{\dagger}} +\cdots \Bigr) \gamma_{\mu}
  W_{\bar{n}}^{\dagger} \xi_{\bar{n}}. 
\end{equation}
The second term in Eq.~(\ref{expxi}) yields the subleading current at
tree level 
\begin{equation} \label{j1a}
  J_{\mu}^{(1a)} = \overline{\xi}_n \frac{\FMslash{\overline{n}}}{2} 
  i\overleftarrow{\FMSlash{D}_n^{\perp}} W_n \frac{1}{\overline{n}\cdot
  \mathcal{P}^{\dagger}} \gamma_{\mu} W_{\bar{n}}^{\dagger}
  \xi_{\bar{n}} = -\overline{n}_{\mu} \overline{\xi}_n 
i\overleftarrow{\FMSlash{D}_n^{\perp}} W_n \frac{1}{\overline{n}\cdot
  \mathcal{P}^{\dagger}}  W_{\bar{n}}^{\dagger}
  \xi_{\bar{n}}. 
\end{equation} 
The second type arises from integrating out the off-shell modes when
the collinear quark $\xi_{\bar{n}}$ emits a collinear gluon
$A_n^{\mu}$, and it is given at tree level as 
\begin{equation} \label{j1b}
  J_{\mu}^{(1b)} = \overline{\xi}_n  \gamma_{\mu} 
  i\FMSlash{D}_n^{\perp} W_n \frac{\FMslash{n}}{2} \frac{1}{n\cdot
  \mathcal{P}} W_{\bar{n}}^{\dagger} \xi_{\bar{n}}
=-n_{\mu} \overline{\xi}_n 
  i\FMSlash{D}_n^{\perp} W_n \frac{1}{n\cdot
  \mathcal{P}} W_{\bar{n}}^{\dagger} \xi_{\bar{n}}.  
\end{equation}
This can be derived by computing the Feynman diagram for the process
and by integrating out the intermediate state with virtuality $p^2
\sim Q^2$. And the result can be made gauge invariant by inserting the
appropriate collinear Wilson lines. A novel
method to derive the operator is the auxiliary field method
\cite{Bauer:2001yt, Bauer:2002nz, Chay:2003ju}.  

There are other subleading current operators involving
$iD_{\bar{n}}^{\perp}$, which can
be obtained by expanding $\xi_{\bar{n}}$ to subleading order and 
by considering the process in which $\xi_n$ emits
$A_{\bar{n}}^{\mu}$. However these subleading operators do not 
contribute to the jet function which is obtained by integrating out
the degrees of freedom of order $p^2 \sim Q\Lambda$ in going down to
$\mathrm{SCET}_{\mathrm{II}}$ because these subleading operators
describe the interaction of the $\overline{n}$-collinear particles in the
proton. These operators contribute to the subleading corrections for
the parton distribution functions which are given by the matrix
elements of the collinear operators in the $\overline{n}^{\mu}$
direction, and we will not consider them here.  

Before going down to $\mathrm{SCET}_{\mathrm{II}}$, it is convenient
to factor out the usoft interactions by redefining the collinear
fields, for example, as
\begin{equation} \label{usoftd}
  \xi_n \rightarrow Y_n \xi_n, \  A_n^{\mu} \rightarrow
  Y_n A_n^{\mu} Y_n^{\dagger}, \ \
\xi_{\bar{n}} \rightarrow Y_{\bar{n}} \xi_{\bar{n}}, \
A_{\bar{n}}^{\mu} \rightarrow Y_{\bar{n}} A_{\bar{n}}^{\mu}
Y_{\bar{n}}^{\dagger},  
\end{equation}
for the collinear fields moving from $-\infty$ to $x$. Once the usoft
interactions are factored out, collinear particles do not interact
with usoft particles any more. The prescription of the usoft Wilson
lines depends on the propagation of the collinear particles or
antiparticles to which the soft gluons are attached, and it is
described in detail in Ref.~\cite{Chay:2004zn}. The possible usoft
Wilson lines are given by  
\begin{eqnarray} \label{usoft}
Y_n &=& \sum_{\rm{perm}} \exp \Biggl[\frac{1}{n\cdot \mathcal{R} +
  i\epsilon} (-g n\cdot A_{us})\Biggr],~~~~ 
Y_n(x) = P \exp\Biggl[ig \int^{x}_{-\infty} ds ~n\cdot A_{us}\Biggr],  
\nonumber \\
\tilde{Y}_n &=& \sum_{\rm{perm}} \exp 
\Biggl[\frac{1}{n\cdot \mathcal{R} - i\epsilon}
(-g n\cdot A_{us})\Biggr],~~~~ 
\tilde{Y}_n(x) = \overline{P} \exp\Biggl[ig \int^{\infty}_{x} ds
~n\cdot A_{us}\Biggr],
\end{eqnarray}    
where $\mathcal{R}$ is the momentum operator for the usoft fields and
the path ordering $P$ means that the fields are ordered such 
that the gauge fields closer (farther) to the point $x$ are moved to
the left, while $\overline{P}$ denotes the anti-path ordering. 
As explained in Ref.~\cite{Chay:2004zn}, $Y_n$ ($Y_n^{\dagger}$) is
the usoft Wilson line attached to the collinear particle
(antiparticle) from $-\infty$, while $\tilde{Y}_n$
($\tilde{Y}_n^{\dagger}$) is the usoft 
line attached to the collinear antiparticle (particle) moving to
$\infty$. This delicate procedure of choosing the appropriate usoft
Wilson lines is related to the $i\epsilon$ prescription, which
specifies the location of the poles. Physically, this is related to
choosing the sign of 
$\overline{n}\cdot p$ since the denominator in Eq.~(\ref{usoft}) is
actually $n\cdot \mathcal{R} +i\mathrm{sgn.} (\overline{n}\cdot
p)\epsilon$, and the sign of $\overline{n}\cdot p$ depends on whether
the collinear field is a particle or an antiparticle.

Now that the current operators at leading and subleading order in SCET
are known, we can compute the forward scattering amplitude $T_{\mu
  \nu}$, or the hadronic tensor $W_{\mu\nu}$ and factorize the usoft
interactions using the appropriate prescription for the usoft Wilson
lines. Then we integrate out the degrees of freedom of order $p^2 \sim
Q\Lambda$ to obtain the result in $\mathrm{SCET}_{\mathrm{II}}$. In
$\mathrm{SCET}_{\mathrm{II}}$ the soft interactions are
decoupled from the collinear particles with $p^2 \sim \Lambda^2$, and
the decoupled soft particles contribute to the soft Wilson lines which
are responsible for the emission of soft gluons. 

\section{Factorization of $F_1 (x,Q)$\label{f1}}
\begin{figure}[t]
\begin{center}
\epsfig{file=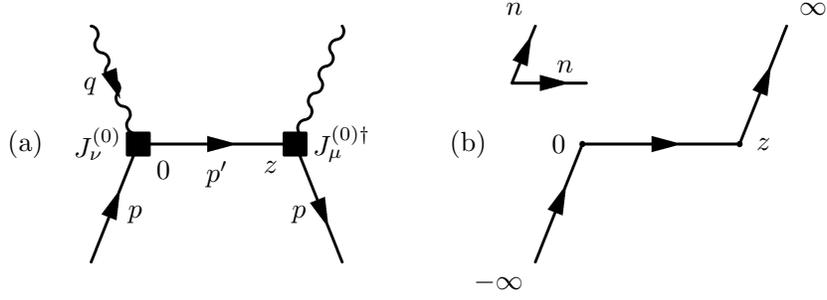, width=11cm}
\end{center}
\vspace{-1.0cm}
\caption{(a) Feynman diagram for the forward scattering amplitude in
DIS in $\rm{SCET_I}$ and (b) the prescription of
the (u)soft Wilson lines.} 
\label{forsc}
\end{figure}

In $\mathrm{SCET}_{\mathrm{I}}$ after the usoft
factorization, the time-ordered product $\hat{T}_{\mu\nu}^{(0)}$ at
leading order is written as 
\begin{eqnarray} \label{tmunu0}
\hat{T}^{(0)}_{\mu\nu} &=& i   C^2(Q) \int d^4 z
  e^{i(q+\tilde{p}-\tilde{p}^{\prime}) 
  \cdot z} T\Bigl[ J_{\mu}^{(0)\dagger} (z) J_{\nu}^{(0)} (0)\Bigr]
\nonumber \\
&=& i   C^2(Q) \int d^4 z
  e^{i(q+\tilde{p}-\tilde{p}^{\prime}) 
  \cdot z} T\Bigl[ \overline{\xi}_{\bar{n}} W_{\bar{n}}
  \tilde{Y}^{\dagger}_{\bar{n}} \gamma_{\mu} Y_n W_n^{\dagger} \xi_n
  (z) \overline{\xi}_n W_n Y_n^{\dagger} \gamma_{\nu} Y_{\bar{n}}
  W_{\bar{n}}^{\dagger} \xi_{\bar{n}} (0) \Bigr], 
\end{eqnarray}
where $\tilde{p}^{\mu}$ and $\tilde{p}^{\prime\mu}$ are the label
momenta. Note that $\overline{n} \cdot q =\overline{n} \cdot
\tilde{p}^{\prime}$ which ensures the conservation of the label
momenta in the $n^{\mu}$ direction, while there is a slight mismatch
in the $\overline{n}^{\mu}$ direction near the endpoint $w\sim 1$ such
that $n\cdot q + n\cdot \tilde{p} = (1-w) n\cdot p$, which survives in
the exponent. The Feynman diagram of the forward scattering
amplitude for $\hat{T}_{\mu\nu}^{(0)}$ is sketched in
Fig.~\ref{forsc} (a).

The prescription for the usoft Wilson lines in DIS  is described in
Fig.~\ref{forsc} (b).  
It is determined by the external states, which consist of an
incoming particle $\xi_{\bar{n}}$ from $-\infty$ to 0, and an outgoing
particle $\overline{\xi}_{\bar{n}}$ from $z$ to $\infty$. The
intermediate states can move either from 0 to $-\infty$ and then from
$-\infty$ to $z$, or from 0 to $\infty$ and then from $\infty$ to
$z$. In both cases, the usoft Wilson line survives between 0 and $z$,
and the remaining part cancels. Either choice of the intermediate
states is appropriate for describing DIS and here we choose the usoft
Wilson lines for each current as 
\begin{eqnarray} \label{usdir}
&& \overline{\xi}_n W_n \gamma_{\mu} W_{\bar{n}}^{\dagger}
 \xi_{\bar{n}}:\  \overline{\xi}_n \rightarrow \overline{\xi}_n
 Y_n^{\dagger}, \ A_n^{\mu} \rightarrow Y_n A_n^{\mu} Y_n^{\dagger}, \
 \xi_{\bar{n}} \rightarrow Y_{\bar{n}} \xi_{\bar{n}}, \
 A_{\bar{n}}^{\mu} \rightarrow Y_{\bar{n}} A_{\bar{n}}^{\mu}
   Y_{\bar{n}}^{\dagger}, \nonumber \\
&& \overline{\xi}_{\bar{n}} W_{\bar{n}} \gamma_{\mu} W_n^{\dagger}
\xi_n: \ \overline{\xi}_{\bar{n}} \rightarrow \overline{\xi}_{\bar{n}}
\tilde{Y}_{\bar{n}}^{\dagger}, \ A_{\bar{n}}^{\mu} \rightarrow
\tilde{Y}_{\bar{n}} A_{\bar{n}}^{\mu} \tilde{Y}_{\bar{n}}^{\dagger}, \
\xi_n \rightarrow Y_n \xi,  \ A_n^{\mu} \rightarrow Y_n A_n^{\mu}
Y_n^{\dagger}, 
\end{eqnarray}
where the intermediate state is going from 0 to $-\infty$, then moving
from $-\infty$ to $z$. This prescription is used in
Eq.~(\ref{tmunu0}).

 Since there are no collinear particles in the $\overline{n}^{\mu}$
 direction in the final state, we obtain the jet function defined by 
\begin{equation}
\langle 0| T[W^{\dagger}_n \xi_n (z) \overline{\xi}_n W_n (0)]
|0\rangle = i\frac{\FMslash{n}}{2} \int \frac{d^4 k}{(2\pi)^4}
e^{-ik\cdot z} J_P (k),   
\end{equation}
where $P$ is the label momentum and $J_P (k)$ depends only on $n\cdot
k$. We can simplify $\hat{T}_{\mu\nu}^{(0)}$ using the fact that 
\begin{equation}
 \int d^4 z \int \frac{d^4 k}{(2\pi)^4} e^{-ik_{\perp}\cdot z_{\perp}
   -i \bar{n} \cdot k n\cdot z/2} = \int d^4 z \frac{1}{4\pi} \int
 dn\cdot k \delta^2 (z_{\perp}) \delta \Bigl( \frac{n\cdot z}{2}
 \Bigr) = \frac{1}{4\pi} \int dn\cdot k d\overline{n} \cdot z, 
\end{equation}
and plugging the jet function into Eq.~(\ref{tmunu0}), we have
\begin{eqnarray} \label{tmunu}
\hat{T}_{\mu\nu}^{(0)}&=& -C^2(Q)
\int d\omega  \int \frac{d\overline{n}\cdot z dn\cdot
  k}{4\pi} e^{i [\omega/2 +n\cdot q-n\cdot  k] \bar{n}\cdot z/2} J_P
  (n\cdot k) \nonumber \\
&\times& T\Bigl[ \overline{\xi}_{\bar{n}} W_{\bar{n}}
  \tilde{Y}_{\bar{n}}^{\dagger}  Y_n
  \Bigl(\frac{\overline{n}\cdot z}{2} \Bigr) \delta (\omega-\mathcal{P}_+)
  \gamma_{\mu}\frac{\FMslash{n}}{2}\gamma_{\nu} 
  Y_n^{\dagger}  Y_{\bar{n}} 
  W_{\bar{n}}^{\dagger} \xi_{\bar{n}} (0) \Bigr] \nonumber \\
&= &- C^2(Q)\int d\omega \int
\frac{d\overline{n}\cdot   z dn\cdot   k}{4\pi} \int d\eta e^{i [
  \omega/2 +n\cdot q -n\cdot  k   -\eta] \bar{n}\cdot   z/2} J_P
(n\cdot k) 
\nonumber \\ 
&\times& T\Bigl[ \overline{\xi}_{\bar{n}}
W_{\bar{n}} \delta (\omega
-\mathcal{P}_+)  \gamma^{\mu}\frac{\FMslash{n}}{2} \gamma_{\nu}
\tilde{Y}_{\bar{n}}^{\dagger} Y_n \delta (\eta 
+n\cdot i\partial) Y_n^{\dagger} Y_{\bar{n}}
W_{\bar{n}}^{\dagger} \xi_{\bar{n}} (0) \Bigr] \nonumber \\
&\longrightarrow& - ~C^2(Q)\int d\omega \int dn\cdot k \int d\eta 
\delta \Bigl( \frac{\omega}{2}+n\cdot q -n\cdot k -\eta\Bigr) J_P
(n\cdot k) \nonumber \\ 
&\times& \frac{1}{N} \langle 0|  \mathrm{tr}\Bigl[
\tilde{S}_{\bar{n}}^{\dagger} S_n \delta (\eta +n\cdot i\partial)
S_n^{\dagger} S_{\bar{n}} \Bigr] |0\rangle  ~\overline{\xi}_{\bar{n}}
W_{\bar{n}} \delta
(\omega - \mathcal{P}_+) \gamma_{\mu} \frac{\FMslash{n}}{2} \gamma_{\nu}
W_{\bar{n}}^{\dagger} \xi_{\bar{n}} (0) \nonumber \\
&=& -C^2 (Q) \int d\omega  \int d\eta J_P \Bigl(
\frac{\omega}{2}+n\cdot q -\eta\Bigr) S(\eta) 
\overline{\xi}_{\bar{n}}
W_{\bar{n}} \delta
(\omega - \mathcal{P}_+) \gamma_{\mu} \frac{\FMslash{n}}{2} \gamma_{\nu}
W_{\bar{n}}^{\dagger} \xi_{\bar{n}},
\end{eqnarray}
where the usoft Wilson line $Y_n$ ($Y_{\bar{n}}$) in
$\mathrm{SCET}_{\mathrm{I}}$ is replaced by the soft Wilson line $S_n$
($S_{\bar{n}}$) in $\mathrm{SCET}_{\mathrm{II}}$. Here the operator
$\mathcal{P}_+ = n\cdot \mathcal{P} + n\cdot \mathcal{P}^{\dagger}$ is
the sum of the label momenta. Since the soft interaction is decoupled
from the collinear sector, the soft Wilson lines are pulled out, and
are described by the vacuum expectation of the soft Wilson line
$S(\eta)$, which is given by
\begin{equation}
  S(\eta) = \frac{1}{N} \langle 0| \mathrm{tr}
 \Bigl[ \tilde{S}_{\bar{n}}^{\dagger} S_n \delta (\eta +n\cdot
 i\partial)   S_n^{\dagger} S_{\bar{n}} \Bigr] |0\rangle. 
\end{equation}
The delta function in Eq.~(\ref{tmunu}) states that the momentum
conservation in the $n$ direction includes the soft momentum from soft
gluons. Eq.~(\ref{tmunu}) is the factorized form for the 
leading forward scattering amplitude. It consists of the hard part
$C^2 (Q)$, obtained in matching the current between the
full theory and $\mathrm{SCET}_{\mathrm{I}}$, the jet function $J_P
(n\cdot k)$, obtained in matching between
$\mathrm{SCET}_{\mathrm{I}}$ and $\mathrm{SCET}_{\mathrm{II}}$, and
the remaining collinear and soft operators in
$\mathrm{SCET}_{\mathrm{II}}$, whose matrix 
elements are given by nonperturbative parameters. The radiative
corrections or the renormalization group evolution of each term can be
computed in perturbation theory. 

Note that the final operators in $\mathrm{SCET}_{\mathrm{II}}$ show a
peculiar structure. In inclusive $B$ decays, the final operator after
the two-step matching is a heavy quark bilinear operator with
the soft Wilson lines. The matrix element of this operator is
parameterized by the shape function of the $B$ meson
\cite{Bauer:2003pi}. This is because the final operator is made of
soft particles. But in DIS, the final operators are made of the
collinear operators and the soft Wilson line. The soft Wilson line is
responsible for the soft gluon emission and the parton distribution
function near the endpoint is affected by this when a collinear
particle participates in the hard scattering. It
is a general feature for physical processes with collinear external
particles that the soft Wilson line does not completely cancel near the
endpoint, and it describes the soft gluon emission in the process.

\section{parton distribution function\label{pdfs}}

We can extract $T_1^{(0)}$, proportional to $-g_{\mu\nu}^{\perp}$ in
$\hat{T}_{\mu\nu}^{(0)}$ from Eq.~(\ref{tmunu}), and it is given by 
\begin{eqnarray} \label{t1zero}
T_1^{(0)} &=& - C^2(Q) \int d\omega 
\int \frac{d z dn\cdot 
  k}{2\pi} \int d\eta  e^{i [\omega/2 +n\cdot q-n\cdot  k -\eta] z} 
J_P  (n\cdot k) S(\eta)\nonumber \\ 
&\times& \langle P|\overline{\xi}_{\bar{n}}
W_{\bar{n}} \delta
(\omega - \mathcal{P}_+) \frac{\FMslash{n}}{2} 
W_{\bar{n}}^{\dagger} \xi_{\bar{n}} (0)|P\rangle_{\mathrm{spin\ av.}},  
\end{eqnarray}
where $z =\overline{n} \cdot z/2$. We want to express
Eq.~(\ref{t1zero}) in terms of the parton distribution functions. Here
we consider only the flavor nonsinglet contribution. The standard 
coordinate space definitions \cite{Soper:1996sn} for the proton parton
distribution functions $f_P^q (y)$ for quarks of flavor $q$ moving in
the $\overline{n}$ direction in full QCD are given as 
\begin{equation} \label{pdfull}
  f_P^q (y) = \int \frac{dz}{2\pi} e^{-iyzn\cdot P} \langle P(P)| 
  \overline{q} (z) Y(z,0) \frac{\FMslash{n}}{2} q
  (0)|P(P)\rangle_{\mathrm{spin\ av.}}, 
\end{equation}
where $Y(y,0)$ is the path-ordered Wilson line and $|P(P)\rangle$ is 
the proton state with momentum $P^{\mu}$. Here $y$ is defined as the
longitudinal momentum fraction of the parton before the hard
scattering and that of the proton, $n\cdot p = y n\cdot P$. 

\begin{figure}[b]
\begin{center}
\epsfig{file=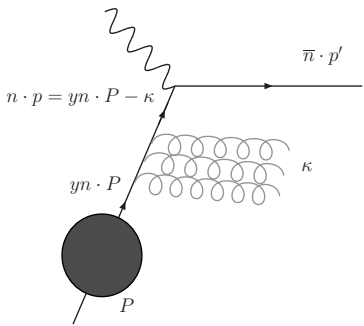, width=5.0cm}
\end{center}
\vspace{-1.0cm}
\caption{An energetic parton comes out of the proton with the momentum
$y n\cdot P$. It emits soft gluons with momentum $\kappa$, and the
momentum of the hard parton before the hard scattering becomes
$yn\cdot P -\kappa$. }
\label{pdscet}
\end{figure}

The definition of the parton distribution function in
Eq.~(\ref{pdfull}) is appropriate away from the endpoint region. But
near the endpoint region, we have to extend the definition of 
the parton distribution to include the effect of the soft gluon
emission, satisfying the requirement that it approach
Eq.~(\ref{pdfull}) away from the endpoint region. At first sight, the
soft momentum does not affect the parton distribution function since
it describes the large energy component of the parton. In
order to see why this is not so, let us consider a parton near the
endpoint region undergoing a hard collision, as depicted in   
Fig.~\ref{pdscet}.  First a parton
with the longitudinal momentum fraction $y$ comes out of the
proton. It emits soft gluons with total momentum $\kappa \sim
\Lambda_{\mathrm{QCD}}$ before it
undergoes a hard collision with a photon. The parton distribution
function $f_P^q(n\cdot p /n\cdot P)$ describes the probability of a
parton entering the hard collision with the longitudinal momentum
fraction $n\cdot p = y n\cdot P -\kappa$. When $yn\cdot P \gg
\Lambda_{\mathrm{QCD}}$ including the endpoint region, the inclusion
of $\kappa$ seems to give a negligible effect. When we take a
time-ordered product of this current as in Fig.~\ref{forsc} (a), all the
soft gluons are attached to the $n$-collinear outgoing fermion due to
the property of the soft interactions, which means that all the soft
gluons are real gluons when we take the discontinuity. Away from the
endpoint region, $n\cdot p^{\prime}$ and $\overline{n}\cdot
p^{\prime}$ of the $n$-collinear quark are of order $Q$, and are not
affected by the interaction of the soft gluons, that is, 
$n \cdot p^{\prime}$ does not change to leading order in
$\Lambda$. Therefore the interaction of the soft gluons can be 
neglected and we can safely put $\kappa =0$.  

Near the endpoint region, however, $n \cdot p^{\prime}$ is
of order $\Lambda$ and the interaction with the soft gluons can
significantly affect $n\cdot p^{\prime}$. If a physical quantity
depends on the term proportional to $1/n\cdot p^{\prime}$, like the
jet function, we have to keep the momentum $\kappa$ of the soft gluons
carefully. From the above argument, the parton distribution function
in Eq.~(\ref{pdfull}) can be extended in the endpoint region as
\begin{equation} \label{pdext}
  f_P^q (y,\kappa) = \int \frac{dz}{2\pi} e^{-iz(yn\cdot P-\kappa)}
  \langle P(P)|   \overline{q} (z) Y(z,0) \frac{\FMslash{n}}{2} q
  (0)|P(P)\rangle_{\mathrm{spin\ av.}}, 
\end{equation}
where the naive $yn\cdot P$ is replaced by $yn\cdot P-\kappa$, that
is, the parton distribution function is a function of the large
longitudinal momentum fraction $y$, and the momentum of the soft
gluons $\kappa$. The Wilson line $Y(z,0)$ in Eq.~(\ref{pdfull}) is the
Wilson line with the gauge field in full QCD. However, by looking at
the kinematics near the endpoint, the Wilson line can be decomposed
into the collinear and the soft Wilson lines
\cite{Korchemsky:1992xv}. This procedure is similar to the approach in
SCET, and SCET makes this procedure manifest.

The parton distribution function in $\mathrm{SCET}_{\mathrm{II}}$ can
be written as
\begin{eqnarray} \label{fscet}
f_P^q (y,\kappa) &=& \int d\omega \int \frac{dz}{2\pi}
e^{i(\omega/2 -   y n \cdot P +\kappa) z} \langle P_{\bar{n}} |
\Bigl[\overline{\xi}_{\bar{n}} W_{\bar{n}}\Bigr] 
\tilde{S}_{\bar{n}}^{\dagger} S_n (z)  
\frac{\FMslash{n}}{2} \delta (\omega -\mathcal{P}_+)
S_n^{\dagger} S_{\bar{n}}
\Bigl[W_{\bar{n}}^{\dagger}  \xi_{\bar{n}} \Bigr] (0) |P_{\bar{n}}
\rangle \nonumber \\
&=& \int d\omega \int \frac{dz}{2\pi}
e^{i(\omega/2 -   y n \cdot P) z} \int d\eta e^{i(\kappa- \eta) z}
\nonumber \\
&&\times S(\eta)  \langle P_{\bar{n}} | \Bigl[\overline{\xi}_{\bar{n}}
W_{\bar{n}}\Bigr] \frac{\FMslash{n}}{2} \delta (\omega -\mathcal{P}_+) 
\Bigl[W_{\bar{n}}^{\dagger}  \xi_{\bar{n}} \Bigr]  |P_{\bar{n}}
\rangle, 
\end{eqnarray}
where the spin average is implied. The
additional exponential factor $e^{i\omega z/2}$ comes from 
the label momentum of $\overline{\xi}_{\bar{n}}$.  The usoft Wilson
lines are prescribed according to Eq.~(\ref{usdir}), and the proton
state $|P\rangle$ is replaced by $|P_{\bar{n}} \rangle$, in which
the valence quarks are collinear in the $\overline{n}$ direction. 

Let us define a new parameter $g_P^q (u)$, given by the
spin-averaged matrix element of the collinear operators, as
\begin{equation}
\langle P_{\bar{n}} | \overline{\xi}_{\bar n} W_{\bar n} \delta
(\omega -\mathcal{P}_+) \frac{\FMslash{n}}{2} W_{\bar n}^{\dagger}
\xi_{\bar n} |P_{\bar n} \rangle = n\cdot P \int du \delta (\omega -2u
n\cdot P) g_P^q (u),
\end{equation}
where the contribution from the antiquark is discarded for
simplicity. Physically $g_P^q (u)$ corresponds to the probability for
the proton to emit a parton with the longitudinal momentum fraction
$u$ before the parton emits soft gluons. Of course, $g_P^q$ is not
physical since we cannot separate a collinear parton from a cloud of
soft gluons. Only after $g_P^q(u)$ is combined with the effect of the
soft gluon emission, the parton distribution $f_P^q$ is physically
meaningful.

The relation between $f_P^q$ and $g_P^q$ is given by
\begin{equation}
  \label{fpgp}
f_P^q (y,\kappa) = \int d\omega \int  \frac{dz}{2\pi}
e^{i(\omega/2 -   y n \cdot P) z} \int d\eta e^{i(\kappa- \eta) z} 
(n\cdot P) S(\eta) \int du \delta (\omega -2un\cdot P)g_P^q (u). 
\end{equation}
Note that $\omega$ and $yn\cdot P$ are the label momenta, and
$\kappa$, $\eta$ are the residual momenta. Therefore using the fact
that
\begin{equation}
  \int dz e^{i(\omega/2 -   y n \cdot P) z} e^{i(\kappa-
    \eta) z}  =\delta_{\omega, yn\cdot P} \int dz e^{i(\kappa-
    \eta) z},
\end{equation}
and integrating the delta function with respect to $\omega$ yield 
\begin{equation} \label{pds}
f_P^q (y,\kappa) = (n\cdot P) S(\kappa) g_P^q (y).   
\end{equation}

In terms of the parton distribution function $f_P^q (y,\kappa)$,
$T_1^{(0)}$ is written as  
\begin{eqnarray}\label{t1two}
T_1^{(0)} &=& -C^2 (Q) \int d\omega \int dn\cdot k \int
\frac{dz}{2\pi} \int d\eta 
e^{i(\omega/2+n\cdot q -n\cdot k -\eta) z} J_P (n\cdot k) (n\cdot P) 
S(\eta)  \nonumber \\ 
&\times& \int du \delta (\omega -2u n\cdot P) g_P^q (u) \nonumber \\
&=&   -C^2 (Q) \int dn\cdot k \int d\eta \int du J_P (n\cdot k)
(n\cdot P) S(\eta) \delta (u n\cdot P +n\cdot q -n\cdot k -\eta) g_P^q
(u)  \nonumber \\
&=&  -C^2 (Q) \int d\eta \int du  J_P (u n\cdot P +n\cdot q
 -\eta) (n\cdot  P) S(\eta) g_P^q (u)\nonumber \\
&=&  -C^2 (Q) \int d\eta  \int du  J_P (u n\cdot P
+n\cdot q   -\eta) f_P^q (u, \eta).
\end{eqnarray}
In deriving this result, note that $u n\cdot P$ is the label momentum
of the parton, that is, $n\cdot p$, and the exponent indicates the
momentum conservation since a slight mismatch between $2u n\cdot P$
and the photon momentum $n\cdot q$ gives $n\cdot k +\eta$. The forward
scattering amplitude is given by a double convolution of the jet
function with the collinear matrix element and the soft Wilson
line. Because the jet function is affected by both the collinear
momentum and the soft momentum, it is impossible to write $T_1^{(0)}$
as a single convolution with the conventional parton distribution
function $f_P^q (y)$ without the effect of the soft gluon
emission. However, as will be shown below, the moment of $f_P^q$ is
given by the product of the moment of the soft Wilson line and $g_P^q$. 

Away from the endpoint region, we can neglect the soft momentum
$\eta$, and in this limit the soft Wilson line cancels to give
$f_P^q (y,0) = g_P^q (y)$ is the conventional parton
distribution function. And we recover the result away from the
endpoint region  
\begin{equation}
 T_1^{(0)} = - \int dy  H(Q,y)  f_P^q (y), 
\end{equation} 
where $H(Q,y)$ is the hard function, which can be split into $C^2 (Q)$
and the jet function near the endpoint region.
The main difference is that the effect of the soft gluon emission
cancels away from the endpoint region, whereas incomplete cancellation
occurs near the endpoint region. This incomplete cancellation results
in the presence of the soft Wilson line, which represents the real
soft gluon emission in the process. 

\section{moment analysis\label{moana}}
In order to consider $T_1^{(0)}$ in moment space, let us introduce
$\eta = (1-v) n\cdot p$. The jet function $J_P(n\cdot k)$ has support
only for a positive argument, and from Eq.~(\ref{t1two}), $n\cdot k$
is given in terms of the partonic variable $n\cdot p$ as
\begin{equation}
  n\cdot k = un\cdot P +n\cdot q -\eta = n\cdot p -w n\cdot p -(1-v)
  n \cdot p = (v-w) n\cdot p. 
\end{equation}
Therefore $v$ should be $w\leq v \leq 1$, and $T_1^{(0)}$ is
written in terms of the partonic variables as
\begin{equation} \label{t1f}
T_1^{(0)} (x,Q) = -  C^2(Q)\int_x^1 dy  \frac{g_P^q (y)}{y} 
\int_w^1 dv (n\cdot p)^2 S\Bigl( (1-v) n\cdot p\Bigr) J_P \Bigl(
(v-w)n\cdot  p\Bigr).
\end{equation}
We take the discontinuity of $T_1^{(0)} (x,Q)$
to obtain the structure function $F_1 (x,Q)$. The hard coefficient
$C(Q)$ and $g_P^q (y)$ are real, therefore the imaginary part arises
from the product of $S\Bigl( (1-v) n\cdot p\Bigr) J_P \Bigl( (v-w)
n\cdot p\Bigr)$. The procedure of taking the discontinuity can be
performed either in $\mathrm{SCET}_{\mathrm{I}}$ or in
$\mathrm{SCET}_{\mathrm{II}}$. Since Eq.~(\ref{t1f}) is the result
obtained in $\mathrm{SCET}_{\mathrm{II}}$, we describe how the 
imaginary part can be taken in $\mathrm{SCET}_{\mathrm{II}}$. 
The discontinuity of $T_1^{(0)}$ 
comes from the jet function $J_P$ only since the soft Wilson line is
real due to the fact that it is hermitian, hence its vacuum
expectation value is real.  The detailed
discussion of taking the imaginary part in
$\mathrm{SCET}_{\mathrm{I}}$ and $\mathrm{SCET}_{\mathrm{II}}$, and
the proof that the imaginary parts in both theories are the same are
presented in Appendix B.

In $\mathrm{SCET}_{\mathrm{II}}$, we obtain the flavor nonsinglet
structure function as  
\begin{equation}
  F_1 (x,Q) =  C^2 (Q) \sum_q e_q^2 \int_x^1 \frac{dy}{y} g_P^q (y)
  \int_w^1 dv  (n\cdot p)^2  S\Bigl( (1-v) n\cdot p\Bigr) \frac{-1}{\pi}
  \mathrm{Im} J_P  \Bigl( (v-w) n\cdot p\Bigr),
\end{equation}
where $e_q$ is the electric charge of the parton $q$. For simplicity,
we omit the summation over the quark flavors $q$ from now on. 
Note that $J_P (n\cdot k)$ is actually the propagator of the
$n$-collinear fermion, so it is of the form $1/(n\cdot k+i0^+)$ modulo
logarithms of $n\cdot k$ with radiative corrections. Let us define the
dimensionless function $\tilde{J}_P$ near the endpoint as
\begin{equation} \label{tildej}
n\cdot p \frac{-1}{\pi} \mathrm{Im} \ J_P ( v-w) = \frac{1}{v}
\tilde{J}_P \Bigl( \frac{w}{v} \Bigr), 
\end{equation}
and let us also define the dimensionless soft Wilson lines as 
\begin{equation}
 \tilde{S} (v) \equiv n\cdot p S\Bigl((1-v) n\cdot p\Bigr)
 = \frac{1}{N} \langle 0| \mathrm{tr}
  \tilde{S}_{\bar{n}}^{\dagger} S_n \delta \Bigl(1-v + \frac{n\cdot
    i\partial}{n\cdot p} \Bigr)   S_n^{\dagger} S_{\bar{n}}
  |0\rangle. 
\end{equation}
Then $F_1 (x,Q)$, with the explicit renormalization
scales, can be written as 
\begin{equation} \label{f1final}
  F_1 (x,Q) =  C^2 (Q,\mu_0)  \int_x^1 \frac{dy}{y} g_P^q (y,\mu) 
\int_w^1 \frac{dv}{v} \tilde{S} (v,\mu) \tilde{J}_P
(\frac{w}{v},\mu_0, \mu),   
\end{equation}
where $\mu_0 \sim Q \sqrt{1-x}$ is the scale between
$\mathrm{SCET}_{\mathrm{I}}$ and $\mathrm{SCET}_{\mathrm{II}}$, and
$\mu$ is the renormalization scale in $\mathrm{SCET}_{\mathrm{II}}$
with $Q\gg\mu_0 \sim Q\sqrt{1-x} \gg \mu$. 
According to the two-step matching, the Wilson coefficient obtained
from the matching at $Q$ evolves down to the scale $\mu_0$. The jet
function is computed from the matching between $\rm{SCET_I}$ and
$\rm{SCET_{II}}$ at $\mu_0$, and it evolves to the scale $\mu$.  
The soft Wilson line and the collinear matrix element are
evaluated at the final scale $\mu$. This is the result in SCET in
comparison to the result obtained in the full QCD factorization approach
\cite{Sterman:1986aj}. 

If we write
\begin{equation}
  B(w) = \int_w^1 \frac{dv}{v} \tilde{S} (v) \tilde{J}_P (\frac{w}{v}), 
\end{equation}
with $w=x/y$, the moment of $F_1 (x,Q)$ can be written as
\begin{eqnarray}
  F_{1,n} &=& C^2 (Q) \int_0^1 dx x^{n-1} \int_x^1 \frac{dy}{y} g_P^q (y)
  B\Bigl( \frac{x}{y}   \Bigr) \nonumber \\
&=&  C^2 (Q) \int_0^1 dx x^{n-1} \int_0^1 dy \int_0^1 dw \delta (x-wy)
g_P^q (y) B(w) \nonumber \\
&=&  C^2 (Q) \int_0^1 dy y^{n-1} g_P^q (y) \int_0^1 dw w^{n-1} B(w)=
g_{P,n}^q \cdot B_n.  
\end{eqnarray}
The $n$-th moment of $B(w)$ can be written as
\begin{eqnarray}
B_n &=& \int_0^1 dw w^{n-1} B(w) 
=\int_0^1 dw w^{n-1} \int_w^1 \frac{dv}{v} \tilde{S} (v) \tilde{J}_P 
 \Bigl(\frac{w}{v} \Bigr) \nonumber \\
&=& \int_0^1 dw w^{n-1} \int_0^1 dz \int_0^1 du \delta (w-uz)
\tilde{S}(z) \tilde{J}_P \Bigl(\frac{w}{v} \Bigr)\nonumber \\
&=& \int_0^1 dz z^{n-1} \tilde{S} (z) \int_0^1 du u^{n-1} \tilde{J}_P
(u) = \tilde{S}_n\cdot \tilde{J}_{P,n}.
\end{eqnarray}
Finally, the moment of the structure function is given as
\begin{equation} \label{f1ns}
  F_{1,n} (Q)=  C^2 (Q,\mu_0) \tilde{J}_{P,n} (\mu_0,\mu)  \cdot
g_{P,n}^q (\mu) \cdot \tilde{S}_n (\mu). 
\end{equation}
The parton distribution function $f_P^q (y,\eta)$ in Eq.~(\ref{pds})
can be written in terms of $y$ and $v$ as 
\begin{equation}
  f_{P}^q (y, v) = (n\cdot p)S\Bigl( (1-v)n\cdot p\Bigr) \frac{g_P^q
    (y)}{y} = \tilde{S} (v) \frac{g_P^q (y)}{y},
\end{equation}
and if we take the double moment of $f_{P}^q (y, v)$, it becomes
\begin{equation}
f_{P,m,n}^q = \int dv v^{m-1} \int dy y^{n-1}  \tilde{S} (v)
\frac{g_P^q (y)}{y} = \tilde{S}_m g_{P,n-1}^q.  
\end{equation}
In terms of the moment $f_{P,m,n}^q$, the moment of the structure
function is given by
\begin{equation}
   F_{1,n} (Q)=  C^2 (Q,\mu_0) \tilde{J}_{P,n} (\mu_0,\mu)  \cdot
   f_{P,n,n+1}^q (\mu).  
\end{equation}
The advantage of SCET in obtaining Eq.~(\ref{f1ns}) is that each
component can be computed independently using perturbation
theory, and we can clearly understand how these terms arise in
SCET.

\section{Radiative corrections\label{radcor}}
We can compute the radiative corrections for each term in the
factorized expression for $F_{1,n}$ in Eq.~(\ref{f1ns}). Let us begin
with the radiative correction for the collinear part $g_P^q$. 
There is a  delicate point in computing the radiative correction of
the collinear part. In any collinear loop integral,
we integrate over all the possible loop momentum and the loop
momentum can reach the region in which collinear particles become
soft. Since the collinear and the soft fields are regarded as distinct
in SCET, we have to remove the soft contribution from the collinear
part to avoid double counting. For this purpose, the zero-bin
subtraction method is suggested \cite{Manohar:2006nz}. Whenever there
is a collinear loop diagram, the loop integration is performed by
counting the loop momentum as collinear. Then the integrand is
rewritten by counting the loop momentum as soft, and the integral
should be subtracted to include only the collinear
contribution. Otherwise, when the soft contribution is included, the
soft contribution is counted twice. This had been missing in SCET and
was first pointed out by Ref.~\cite{Manohar:2006nz}. Some 
previous calculations are not affected by the zero-bin subtraction,
but conceptually the zero-bin subtraction is the correct step to avoid 
double counting. DIS is one of the examples in which the zero-bin
subtraction should be performed carefully.

\begin{figure}[t]
\begin{center}
\epsfig{file=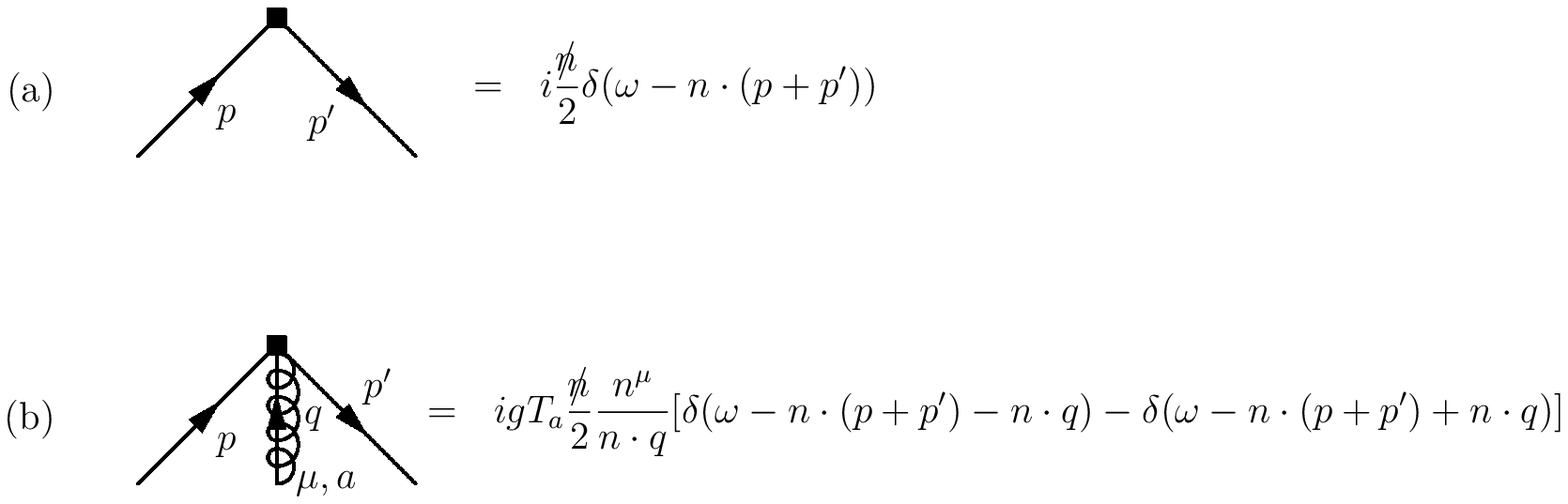, width=14.0cm}
\end{center}
\vspace{-0.5cm}
\caption{Feynman rules for the collinear operator $O_c^q$. (a)
the tree-level operator (b) the operator with a
collinear gluon with incoming momentum $q^{\mu}$.}
\label{ocqf}
\end{figure}

\begin{figure}[b]
\begin{center} \hspace{-2.0cm}
\epsfig{file=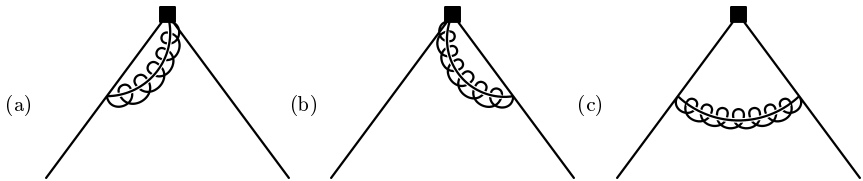, width=8.0cm}
\end{center}
\vspace{-1.2cm}
\caption{Radiative corrections for the quark distribution operator at
  one loop.}
\label{ocqrad}
\end{figure}

Let us define the collinear operator $O_c^q$ ,the
matrix element of which yields $g_P^q (y)$, as
\begin{equation}
  O_q^c =\overline{\xi}_{\bar{n}} W_{\bar{n}} \frac{\FMslash{n}}{2}
  \delta (\omega -\mathcal{P}_+) W_{\bar{n}}^{\dagger} \xi_{\bar{n}}. 
\end{equation}
The Feynman rules for $O_c^q$ including a single gluon are shown in
Fig.~\ref{ocqf}.
And the Feynman diagrams for the radiative corrections of  $O_c^q$
at one loop are shown in Fig.~\ref{ocqrad}. 
The naive radiative corrections without the zero-bin subtraction,
using the dimensional regularization with $D=4-2\epsilon$, are given
as 
\begin{eqnarray} \label{mabc}
  M_a &=& M_b
= -2ig^2 C_F \frac{\FMslash{n}}{2} \int \frac{d^D l}{(2\pi)^D}
\frac{n\cdot (l+p)}{l^2 (l+p)^2 n\cdot l} \Bigl[ \delta (\omega
-\omega^{\prime}) -\delta (\omega -\omega^{\prime} -2 n\cdot l)\Bigr]
\nonumber \\
&=&\frac{\alpha_s C_F}{4\pi} \frac{\FMslash{n}}{2}
  \frac{2}{\epsilon} \Bigl[ \delta (\omega -\omega^{\prime})
  +\frac{\omega}{\omega^{\prime}} \Bigl( \frac{\theta (\omega) \theta
    (\omega^{\prime} -\omega)}{(\omega^{\prime} -\omega)_+}
  +\frac{\theta (-\omega)\theta (\omega-\omega^{\prime})}{(\omega -
    \omega^{\prime})_+}\Bigr) \Bigr], \nonumber \\
M_c &=& -ig^2 C_F  \frac{\FMslash{n}}{2} \int \frac{d^D l}{(2\pi)^D}
\frac{(D-2)l_{\perp}^2 }{l^2 [(l+p)^2]^2} \delta (\omega
-\omega^{\prime} -2n\cdot l) \nonumber \\
&=&\frac{\alpha_s C_F}{4\pi} \frac{\FMslash{n}}{2}
  \frac{1}{\epsilon} \frac{2 (\omega^{\prime}
    -\omega)}{(\omega^{\prime})^2} \Bigl[ \theta (\omega) \theta
  (\omega^{\prime} -\omega) +\theta (-\omega) \theta (\omega
  -\omega^{\prime}) \Bigr], 
\end{eqnarray}
where $\omega^{\prime} = n\cdot (p+p^{\prime})$. Note that the terms
proportional to $\theta (\omega)$ ($\omega >0$) in Eq.~(\ref{mabc})
contribute to the quark distribution function, while those with
$\omega<0$ contribute to the antiquark distribution
function. Therefore the sum of all the corrections contributing to
the quark distribution function is given by
\begin{equation} \label{qdis}
  \Bigl[M_a +M_b + M_c\Bigr]_q = \frac{\alpha_s C_F}{2\pi}
  \frac{\FMslash{n}}{2} 
  \frac{1}{\epsilon} \Bigl[ 2\delta (\omega -\omega^{\prime})
  +\frac{1+(\omega/\omega^{\prime})^2}{(\omega^{\prime} -\omega)_+}
  \theta (\omega^{\prime} -\omega) \theta (\omega) \Bigr], 
\end{equation}
while the contribution to the antiquark distribution function is
obtained by replacing $\omega$ and $\omega^{\prime}$ by $-\omega$ and
$-\omega^{\prime}$ respectively in Eq.~(\ref{qdis}).

The zero-bin contribution in each diagram is obtained by the loop
integral in Eq.~(\ref{mabc}), where the collinear loop momentum covers
the soft region in which $n\cdot l \sim \Lambda$ and
$\overline{n}\cdot l \sim \Lambda^2$\begin{eqnarray}
M_a^0 &=& M_b^0 = -2ig^2 C_F \frac{\FMslash{n}}{2}\int \frac{d^D
  l}{(2\pi)^D} 
\frac{1}{l^2 (\overline{n} \cdot l +p^2/n\cdot p) n\cdot l} \Bigl[
\delta (\omega -\omega^{\prime}) -\delta (\omega -\omega^{\prime} -2
n\cdot l)\Bigr], \nonumber \\
M_c^0 &=&  -ig^2 C_F  \frac{\FMslash{n}}{2} \frac{1}{(n\cdot p)^2}
\int \frac{d^D l}{(2\pi)^D} \frac{(D-2)l_{\perp}^2 }{l^2 [\overline{n}
  \cdot l +p^2/n\cdot p]^2} \delta (\omega -\omega^{\prime} -2n\cdot
l). 
\end{eqnarray}
Here $M_c^0$ is suppressed by $\Lambda^2/Q^2$ and it becomes zero when
performing the loop integration. The total zero-bin contribution
becomes  
\begin{eqnarray}
M_a^0 +M_b^0 +M_c^0 &=& - \frac{\FMslash{n}}{2} \frac{\alpha_s
  C_F}{\pi} \Bigl[ \Bigl( \frac{1}{\epsilon_{\mathrm{UV}}}
-\frac{1}{\epsilon_{\mathrm{IR}}} \Bigr)
\Bigl(\frac{1}{\epsilon_{\mathrm{UV}}} -\ln \frac{-p^2}{\mu n\cdot
  p}\Bigr) \delta(\omega-\omega^{\prime}) \nonumber \\
&-&\frac{1}{\epsilon_{\mathrm{UV}}} \Bigl(\frac{-p^2}{\mu n\cdot p}
  \Bigr)^{-\epsilon} 
\Bigl(\frac{\omega^{\prime} - \omega}{2} \Bigr)^{-1-\epsilon} \theta
(\omega^{\prime} -\omega) \Bigr]. 
\end{eqnarray}
Since
\begin{equation}
\int_{-\infty}^{\infty} d\eta \eta^{-1-\epsilon} \theta(\eta) =
\int_0^{\infty} d\eta \eta^{-1-\epsilon} =
\frac{1}{\epsilon_{\mathrm{UV}}} -\frac{1}{\epsilon_{\mathrm{IR}}},    
\end{equation}
we can write
\begin{equation}
\eta^{-1-\epsilon} \theta (\eta) = \Bigl(
\frac{1}{\epsilon_{\mathrm{UV}}} -\frac{1}{\epsilon_{\mathrm{IR}}}
\Bigr) \delta (\eta) + \frac{\theta(\eta)}{\eta_+},   
\end{equation}
where the subscript means the ``+'' distribution. Using this relation,
the terms proportional to $\delta (\omega-\omega^{\prime})$ cancel,
and the divergent part of the zero-bin contribution is written as
\begin{equation}\label{zfq}
\frac{\FMslash{n}}{2} \frac{\alpha_s C_F}{\pi}
\frac{1}{\epsilon_{\mathrm{UV}}} \frac{\theta(\omega^{\prime}
  -\omega)}{(\omega^{\prime} -\omega)_+}.  
\end{equation}
As will be shown below, this is exactly the same as the soft
contribution from the radiative corrections for $S(\eta)$, and it
should be subtracted from Eq.~(\ref{qdis}).

The relation between the bare operator
$O_q^{cB}$ and the renormalized operator $O_q^{cR}$ can be written as
\begin{equation}
 O_q^{cB} (\omega) =\int d\omega^{\prime} Z(\omega, \omega^{\prime})
   O_q^{cR} (\omega^{\prime}), 
\end{equation}
where the counterterm $Z(\omega, \omega^{\prime})$ is given by
\begin{equation}
  Z(\omega,\omega^{\prime}) = \delta (\omega -\omega^{\prime})
  +\frac{\alpha_s C_F}{2\pi \epsilon} \Bigl[ \frac{3}{2}\delta (\omega
  -\omega^{\prime}) +
  \frac{-1+(\omega/\omega^{\prime})^2}{(\omega^{\prime} -\omega)_+} 
  \theta (\omega^{\prime} -\omega) \theta (\omega) \Bigr].
\end{equation}
The renormalization group equation for $O_q^{cR}$ is given by
\begin{equation} \label{renoq}
  \mu\frac{d}{d\mu} O_q^{cR} (\omega) = -\int d\omega^{\prime} \gamma
  (\omega, \omega^{\prime}) O_q^{cR} (\omega^{\prime}),
\end{equation}
where the anomalous dimension $\gamma (\omega,\omega^{\prime})$ is
given by
\begin{equation}
\gamma (\omega,\omega^{\prime}) = - \frac{\alpha_s C_F}{\pi}
\Bigl[ \frac{3}{2}\delta (\omega   -\omega^{\prime}) +
  \frac{-1+(\omega/\omega^{\prime})^2}{(\omega^{\prime} -\omega)_+} 
  \theta (\omega^{\prime} -\omega) \theta (\omega) \Bigr].
\end{equation}

In order to express Eq.~(\ref{renoq}) in terms of dimensionless
variables, let us write $\omega = 2Ey$, $\omega^{\prime} =2Ez$, where
$E$ is the energy of the quark and $0<y,z<1$. The renormalization
group equation Eq.~(\ref{renoq}) is written as
\begin{eqnarray}\label{renq}
 \mu\frac{d}{d\mu} O_q^{cR} (\omega) &=& \frac{\alpha_s C_F}{\pi} \int
 \frac{dz}{z} \Bigl[ \frac{3}{2} \delta \Bigl( 1-\frac{y}{z} \Bigr)
 +\frac{-1+(y/z)^2}{(1-y/z)_+} \Bigr] O_q^{cR} (z) \nonumber \\
&=&\frac{\alpha_s}{\pi} \int_y^1 \frac{dx}{x} \Bigl[P_{qq} (x)
-\frac{2C_F}{(1-x)_+} \Bigr] O_q^{cR}
\Bigl(\frac{y}{x} \Bigr),
\end{eqnarray}
where $P_{qq} (x)$ is the quark splitting function 
\begin{equation}
  P_{qq} (x) =C_F \Bigl[ \frac{3}{2} \delta (1-x)
  +\frac{1+x^2}{(1-x)_+} \Bigr]. 
\end{equation}
Note that, in Eq.~(\ref{renq}), there is an additional term
$-2C_F/(1-x)_+$ due to the zero-bin subtraction. Therefore the matrix
element $g_P^q$ of $O_q^c$ scales differently from the conventional
parton distribution function $f_P^q (y)$ away from the endpoint
region. However, when we include the effects of the soft Wilson line,
we obtain the same result as the conventional approach. (See below.)  
The moment $g_{P,n}^q$ satisfies the renormalization group equation 
\begin{equation}
\mu \frac{d}{d\mu} g_{P,n}^q = -\gamma_C \cdot g_{P,n}^q,    
\end{equation}
where the anomalous dimension $\gamma_C$ is given as
\begin{equation} \label{gamq}
 \gamma_C =-\frac{\alpha_s}{\pi} \int_0^1 dx x^{n-1} \Bigl[ P_{qq}
 (x) -\frac{2C_F}{(1-x)_+} \Bigr] \rightarrow 
 \frac{\alpha_s C_F}{2\pi} \Bigl[ (4\ln \overline{N} -3) -4\ln
 \overline{N}
 \Bigr] = -3 \frac{\alpha_s C_F}{2\pi}.
\end{equation}
The last expression is the large $n$ limit with $\overline{N} =
ne^{\gamma_E}$. The first parenthesis comes from the splitting
function $P_{qq}(x)$ and the second parenthesis comes from the
zero-bin subtraction $-2C_F/(1-x)_+$.

We now turn to the radiative correction for the soft Wilson line 
$S(\eta)$. The radiative correction for the soft Wilson line was
computed in Ref.~\cite{Chay:2004zn}, and we quote the result. The
relation between the bare operator $S_B (\eta)$ and the renormalized
operator $S_R (\eta)$ is given by 
\begin{equation}
  S_B (\eta) =\int d\eta^{\prime} Z^S_{\mathrm{DIS}} (\eta,
  \eta^{\prime}) S_R (\eta^{\prime} ), 
\end{equation}
where 
\begin{equation} 
  Z^S_{\mathrm{DIS}} (\eta, \eta^{\prime}) = \delta (\eta
  -\eta^{\prime}) +\frac{\alpha_s C_F}{\pi} \frac{1}{\epsilon}
  \frac{\theta (\eta-\eta^{\prime})}{(\eta-\eta^{\prime})_+}. 
\end{equation}
The renormalization group equation for the dimensionless
$\tilde{S}(v)$ is obtained by putting $\eta = (1-v) n\cdot p$,
$\eta^{\prime} = (1-v^{\prime}) n\cdot p$, and it is given as
\begin{equation}
  \mu\frac{d}{d\mu} \tilde{S} (v) = \frac{2\alpha_s C_F}{\pi}
 \int_v^1 \frac{dv^{\prime}}{v^{\prime}} \frac{\tilde{S}
    (v^{\prime})}{\Bigl( 1-v/v^{\prime}\Bigr)_+}   = \frac{2\alpha_s
    C_F}{\pi} \int_v^1 \frac{dv^{\prime}}{v^{\prime}} \frac{\tilde{S}
    (v/v^{\prime})}{( 1-v^{\prime})_+}. 
\end{equation}
The $n$-th moment of the soft Wilson line $\tilde{S} (v)$ satisfies
the renormalization group equation
\begin{equation}
  \mu \frac{d}{d\mu} \tilde{S}_n  = -\gamma_S \cdot \tilde{S}_n, 
\end{equation}
and the anomalous dimension $\gamma_S$ is given as
\begin{equation} \label{gams}
  \gamma_S = \frac{2\alpha_s C_F}{\pi} H_{n-1} \rightarrow
  \frac{\alpha_s C_F}{2\pi} 4\ln \overline{N}, 
\end{equation}
where $H_n = \sum_{j=1}^n 1/j$, and the large $n$ limit is taken in
the final expression. 

Note that $\gamma_S$ is exactly the zero-bin contribution, as can be
seen in Eq.~(\ref{gamq}). Then the double moment $f_{P,n,n+1}^q = 
\tilde{S}_n g_{P,n}^q$ satisfies the renormalization equation
\begin{equation}
\mu \frac{d}{d\mu} f_{P,n,n+1}^q = -\gamma_n  f_{P,n,n+1}^q,   
\end{equation}
where the anomalous dimension
\begin{equation} \label{gamman}
\gamma_n = \gamma_C +\gamma_S = \frac{\alpha_s C_F}{2\pi} \Bigl[ 1 -
\frac{2}{n(n+1)}   +4\sum_{j=2}^n \frac{1}{j}\Bigr] \rightarrow
\frac{\alpha_s C_F}{2\pi} (4 \ln \overline{N} -3) 
\end{equation}
is the one obtained from the Altarelli-Parisi kernel. Therefore 
in moment space, the (double) moment of the parton distribution
function even in the endpoint region satisfies the same
renormalization group equation away from the endpoint region, or the
result in the full theory. 

Finally, let us consider the radiative corrections to the jet
function. To order $\alpha_s$, the jet function $J_P (n\cdot k)$ is
given as 
\begin{equation}
J_P (n\cdot k) = \frac{1}{n\cdot k +i\epsilon} \Bigl[ 1+\frac{\alpha_s
  C_F}{4\pi} \Bigl( 2\ln^2 \frac{-Pn\cdot k-i\epsilon}{\mu^2} -3 \ln
  \frac{-Pn\cdot k-i\epsilon}{\mu^2} + 7-\frac{\pi^2}{3} \Bigr) \Bigr],
\end{equation}
where $P$ is the label momentum $P =\overline{n}\cdot
p^{\prime}=Q$. Therefore $J_Q \Bigl((v-w) n\cdot p \Bigr)$ is written
as 
\begin{eqnarray}
  J_Q \Bigl( (v-w) n\cdot p\Bigr) &=& \frac{1}{(v-w) n\cdot p
    +i\epsilon}  \Bigl[ 1 +\frac{\alpha_s C_F}{4\pi} \Bigl( 2\ln^2
  \frac{-(v-w) Qn\cdot p -i\epsilon}{\mu^2} \nonumber \\
&&-3\ln  \frac{-(v-w)
    Qn\cdot p -i\epsilon}{\mu^2} +7-\frac{\pi^2}{3} \Bigr) \Bigr].  
\end{eqnarray}
The imaginary part of $J_Q\Bigl((v-w) n\cdot p \Bigr)$ is given by
\begin{eqnarray}
\frac{-n\cdot p}{\pi} \mathrm{Im} J_Q  \Bigl((v-w) n\cdot p\Bigr) 
&=& \delta (v-w) \Bigl[ 1+\frac{\alpha_s C_F}{4\pi} \Bigl( 2\ln^2
\frac{Qn\cdot p}{\mu^2} -3 \ln \frac{Qn\cdot p}{\mu^2} +7-\pi^2 \Bigr)
\Bigr] \nonumber \\
&+& \frac{\alpha_s C_F}{4\pi} \Bigl[ \frac{1}{(v-w)_+} \Bigl( 4\ln
\frac{Qn\cdot p}{\mu^2} -3\Bigr) +\frac{4\ln
  (v-w)}{(v-w)_+} \Bigr]. 
\end{eqnarray}

The dimensionless jet function  $\tilde{J}_Q (w/v)$, defined in
Eq.~(\ref{tildej}), with $n\cdot p = Q/w$ can be written as
\begin{eqnarray}
  \tilde{J}_Q \Bigl( \frac{w}{v} \Bigr) &=& \delta \Bigl(
  1-\frac{w}{v} \Bigr) \Bigl[ 1+\frac{\alpha_s C_F}{4\pi} \Bigl(2
  \ln^2 \frac{Q^2}{\mu^2} -3\ln  \frac{Q^2}{\mu^2} +7-\pi^2\Bigr)
  \Bigr] \nonumber \\
&+& \frac{\alpha_s C_F}{4\pi} \frac{1}{(1-w/v)_+} \Bigl( 4 \ln
 \frac{Q^2}{\mu^2} -3+ 4\ln (1-w/v) \Bigr),
\end{eqnarray}
where we neglect the $\ln v$ term as $v\rightarrow 1$. The moment of
$\tilde{J}_Q$ is given as
\begin{eqnarray}
 \tilde{J}_{Q,n} &=& \int_0^1 du u^{n-1} \tilde{J}_Q (u) \nonumber
  \\
&=& 1+\frac{\alpha_s C_F}{4\pi} \Bigl[ 2\ln^2 \frac{Q^2}{\mu^2} -3 \ln
\frac{Q^2}{\mu^2} +7-\pi^2 
+ \Bigl( 4\ln \frac{Q^2}{\mu^2} -3 \Bigr)H_{n-1}
- \sum_{k=1}^{n-1} \frac{4H_k}{k}  \Bigr]
\nonumber \\
&\rightarrow& 1 +\frac{\alpha_s C_F}{4\pi} \Bigl( 2\ln^2
\frac{Q^2}{\overline{N} \mu^2} -3 \ln \frac{Q^2}{\overline{N} \mu^2}
+7-\frac{2\pi^2}{3} \Bigr), 
\end{eqnarray}
where the last expression is obtained in the large $n$ limit, which is
consistent with the result in Ref.~\cite{Manohar:2003vb}. 

We can present the moment of $F_1 (x,Q)$ to order $\alpha_s$. It is
the product of the square of the hard coefficient $C(Q)$, twice the
running of the hard coefficient from $Q$ to $Q/\sqrt{\overline{N}}$
using Eq.~(\ref{evc}), the jet function at $Q/\sqrt{\overline{N}}$,
and the running from $Q/\sqrt{\overline{N}}$ to $\mu$ using
Eqs.~(\ref{gamq}), (\ref{gams}): 
\begin{eqnarray} \label{fal}
  F_{1,n} (Q) &=& C^2 (Q,Q/\sqrt{\overline{N}}) \tilde{J}_{P,n}
  (Q/\sqrt{\overline{N}}, \mu) \tilde{S}_n (\mu) g_{P,n}^q (\mu) 
  \nonumber \\
&=& C^2 (Q) e^{2\gamma_H \ln[(Q/\sqrt{\overline{N}})/Q]}
\tilde{J}_{P,n} (Q/\sqrt{\overline{N}})e^{(\gamma_C +\gamma_S) \ln
  \mu/(Q/\sqrt{\overline{N}})} \tilde{S}_n (\mu) g_{P,n}^q (\mu) 
\nonumber \\
&=& \Biggl[ 1+\frac{\alpha_s C_F}{4\pi} \Bigl( -16+\frac{\pi^2}{3}
\Bigr) +\frac{\alpha_s C_F}{4\pi} \Bigl( -2\ln^2 \overline{N} +6\ln
\overline{N} \Bigr) \nonumber \\
&&+\frac{\alpha_s C_F}{4\pi} \Bigl( 7-\frac{2\pi^2}{3} \Bigr)
+\frac{1}{2} (\gamma_C +\gamma_S) \ln \frac{\mu^2}{Q^2/\overline{N}}
\Biggr] \tilde{S}_n (\mu) g_{P,n}^q (\mu) 
\nonumber \\
&=& \Biggl[ 1+\frac{\alpha_s C_F}{4\pi} \Bigl( 2\ln^2 \overline{N}
+3\ln \overline{N}-\frac{\pi^2}{3} -9\Bigr) +(\gamma_C +\gamma_S) \ln
\frac{\mu}{Q}\Biggr] \tilde{S}_n (\mu)  g_{P,n}^q (\mu).  
\end{eqnarray}
This result can be compared to the DIS structure
function to one loop in Ref.~\cite{Bardeen:1978yd}, where the moments
of the nonsinglet structure function $F_2/(2x)=F_1$ are given as 
\begin{equation} \label{mn}
  M_N = \Bigl[ 1+\frac{\alpha_s}{4\pi} B_{2,N}^{\mathrm{NS}} +\gamma_q
  \ln\frac{\mu}{Q}\Bigr] A_N (\mu). 
\end{equation}
Here $A_N (\mu)$ are the matrix elements of the twist-two operators
renormalized at $\mu$, and $\gamma_q$ is equal to $\gamma_n = \gamma_C
+\gamma_S$, given in Eq.~(\ref{gamman}). In
the large $N$ limit, $B_{2,N}^{\mathrm{NS}}$ is given by
\begin{equation}
  B_{2,N}^{\mathrm{NS}} \rightarrow C_F \Bigl[ 2\ln^2 \overline{N}
  +3\ln \overline{N} -\frac{\pi^2}{3} -9\Bigr]. 
\end{equation}
Eqs.~(\ref{fal}) and (\ref{mn}) are the same, which means that the
result for the  moment of the structure function in the full theory
away from the endpoint region can be extended to the endpoint
region. However, our result is obtained near the endpoint region where
the effect of the soft gluon emission from the soft Wilson line is
present. The fact that the full-theory result can be extended to the
endpoint region results from the relation $f_{P,n,n+1}^q = \tilde{S}_n 
g_{P,n}^q$ in moment space. Here the soft gluon emission plays an
important role, and the effect shows up not only in DIS, but also in
other high-energy processes such as Drell-Yan processes and jet
production in $e^+ e^-$ collisions. In the full theory it was
considered in Ref.~\cite{Catani:1989ne}.    

Eq.~(\ref{fal}) is the result at order $\alpha_s$. We can use the
renormalization group equation to sum up all the large logarithms. 
The moments of the structure function in SCET to leading logarithmic
accuracy is given by 
\begin{equation} 
F_{1,n}(Q) = C^2(Q) e^{-2I_1\Bigl(Q,Q/\sqrt{\overline{N}}\Bigr)} 
\tilde{J}_{P,n}\Bigl(\frac{Q}{\sqrt{\overline{N}}}\Bigr)
e^{-I_2\Bigl(Q/\sqrt{\overline{N}}, \mu\Bigr)} 
\tilde{S}_n (\mu) g_{P,n}^q (\mu), 
\label{expo} 
\end{equation} 
where 
\begin{equation} \label{I12}
I_1 \Bigl(Q, \frac{Q}{\sqrt{\overline{N}}}\Bigr)
= \int^{Q}_{ Q/\sqrt{\overline{N}}}
\frac{d\mu^{\prime}}{\mu^{\prime}} \gamma_H (\mu^{\prime}), \ \ 
I_2 \Bigl(\frac{Q}{\sqrt{\overline{N}}}, \mu \Bigr)
= \int^{ Q/\sqrt{\overline{N}}}_{\mu} 
\frac{d\mu^{\prime}}{\mu^{\prime}} \Bigl( \gamma_C + \gamma_S \Bigl)
(\mu^{\prime}).  
\end{equation}
When we resum the large logarithms in Eq.~(\ref{I12}), it is written
as  
\begin{eqnarray} 
C^2 (Q) e^{-2I_1} 
&=& C^2 (Q) (\overline{N})^{-4C_F/\beta_0}
\Biggl[\frac{\alpha_s (Q/\sqrt{\overline{N}})}{\alpha_s (Q)}\Biggr]^{2C_F 
(3-8\pi/(\beta_0 \alpha_s (Q)))/\beta_0},
\nonumber \\ 
\label{solution} 
\tilde{J}_{P,n} \Bigl(Q/\sqrt{\overline{N}}\Bigr) e^{-I_2} &=&
\tilde{J}_{P,n} \Bigl(Q/\sqrt{\overline{N}}\Bigr) 
\Biggl[\frac{\alpha_s (Q/\sqrt{\overline{N}})}{\alpha_s 
(\mu)}\Biggr]^{-C_F (8\ln \overline{N} -3)/\beta_0}, 
\end{eqnarray} 
with $\beta_0 =11-2n_f/3$. Eq.~(\ref{fal}) is obtained by expanding
Eq.~(\ref{expo}) to first order in $\alpha_s$.

\section{Factorization of $F_L (x,Q)$\label{fl}}
The longitudinal structure function $F_L (x,Q)$ is proportional to
the imaginary part of $T_2 (x,Q)$, as in Eq.~(\ref{stt}). 
If we consider the tensor structure of the
time-ordered products of the currents in SCET, $T_2 (x,Q)$ is obtained
by the products of the subleading currents $ J_{\mu}^{(1a)}$ and
$J_{\mu}^{(1b)}$. Since the tree-level amplitudes vanish, we consider
the operators with $n$-collinear gluons from Eqs.~(\ref{j1a}) and
(\ref{j1b})  
\begin{eqnarray} \label{j1ab}
 J_{\mu}^{(1a)} &\rightarrow& -\overline{n}_{\mu} \int d\omega
 B_a(\omega) 
 \Bigl[\overline{\xi}_n W_n \delta (\omega -\overline{n} \cdot
 \mathcal{P}^{\dagger}) \Bigr] \Bigl[ W_n^{\dagger}  i
 \overleftarrow{\FMSlash{D}_n^{\perp}} W_n\Bigr]  
 \frac{1}{\overline{n} \cdot 
   \mathcal{P}^{\dagger}} W_{\bar{n}}^{\dagger} \xi_{\bar{n}} =
 -\overline{n}_{\mu} j^{(1a)},
 \nonumber \\  
J_{\mu}^{(1b)} &\rightarrow& -n_{\mu} \int d\omega B_b (\omega)
\Bigl[\overline{\xi}_n W_n \delta  (\omega -\overline{n} \cdot
 \mathcal{P}^{\dagger}) \Bigr] \Bigl[W_n^{\dagger}
 i\FMSlash{D}_n^{\perp} W_n \Bigr]
\frac{1}{n\cdot \mathcal{P}} W_{\bar{n}}^{\dagger} \xi_{\bar{n}}
=-n_{\mu} j^{(1b)}.
\end{eqnarray}
Here the delta functions are
included for convenience, and $B_a$, $B_b$ are the Wilson coefficients
for the subleading current operators. 
As explained in Section II, the possible four types of the time-ordered
products with $J_{\mu}^{(1a)}$ and $J_{\mu}^{(1b)}$ should contribute
in the same way due to current conservation. Here we choose the two
possible time-ordered products $T[J_{\mu}^{(1a)} (z) J_{\nu}^{(1a)}
(0)]$ and $T[J_{\mu}^{(1b)} (z) J_{\nu}^{(1b)} (0)]$ and verify that
both contributions are the same by explicit calculation. And we show
that the expression for the longitudinal structure function also
factorizes.

\begin{figure}[t]
\begin{center}
\epsfig{file=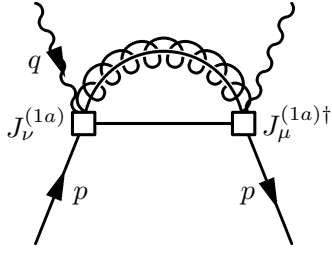, width=4.3cm}
\end{center}
\vspace{-1.0cm}
\caption{The Feynman diagram describing the forward scattering
  amplitude $T_2 (x,Q)$. } 
\label{figt2}
\end{figure}

From the time-ordered product with $J_{\mu}^{(1a)} (z)$ and
$J_{\nu}^{(1a)} (0)$, $\hat{T}_2(x,Q)$ is written as 
\begin{eqnarray} \label{taa}
\hat{T}_2^{aa} (x,Q) &=& 
i\int d^4 z e^{i (q-\tilde{p}^{\prime}+\tilde{p})\cdot z} 
\int d\omega d\omega^{\prime}d\w ~B_a(\omega,\w) 
B_a (\omega^{\prime},\w)  
 \nonumber \\
&\times& T\Biggl[ \overline{\xi}_{\bar{n}} W_{\bar{n}}
\frac{1}{\overline{n} \cdot \mathcal{P}} \Bigl[ W_n^{\dagger}
i\FMSlash{D}_n^{\perp} W_n \Bigr] \Bigl[ \delta (\omega - 
\overline{n} \cdot \mathcal{P}) W_n^{\dagger} \xi_n \Bigr] (z)
\nonumber \\
&\times&
\Bigl[ \overline{\xi}_n W_n \delta (\omega^{\prime} -\overline{n}
\cdot \mathcal{P}^{\dagger} ) \Bigr] \Bigl[W_n^{\dagger}
i\overleftarrow{\FMSlash{D}_n^{\perp}} W_n \Bigr]
\frac{1}{\overline{n} \cdot   \mathcal{P}^{\dagger}}
\delta(\w- \mathcal{P}_+) 
W_{\bar{n}}^{\dagger} \xi_{\bar{n}} (0)\Biggr] \nonumber \\
&\longrightarrow& i \int d\omega d\w \int d^4 z e^{i(1-w)(\w/2)
  \bar{n}\cdot z/2}  B_a^2(\omega,\w) 
 \nonumber \\
&\times& T\Biggl[ \overline{\xi}_{\bar{n}} W_{\bar{n}}
\tilde{Y}_{\bar{n}}^{\dagger} \frac{1}{\overline{n} \cdot \mathcal{P}}
Y_n \Bigl[ W_n^{\dagger} i\FMSlash{D}_n^{\perp} W_n \Bigr]  \Bigl[
\delta (\omega -\overline{n} \cdot \mathcal{P}) W_n^{\dagger}
\xi_n\Bigr] (z) \nonumber \\
&\times& \Bigl[\overline{\xi}_n W_n \Bigr] \Bigl[ W_n^{\dagger}
i\overleftarrow{\FMSlash{D}_n^{\perp}} W_n \Bigr] Y_n^{\dagger}
\frac{1}{\overline{n} \cdot \mathcal{P}^{\dagger}} Y_{\bar{n}}
\delta(\w- \mathcal{P}_+) W_{\bar{n}}^{\dagger} \xi_{\bar{n}} (0) \Biggr], 
\end{eqnarray}
where the last expression is obtained by factorizing the usoft
interactions after redefining the collinear fields. The Feynman
diagram for $T_2 (x,Q)$ is schematically shown in Fig.~\ref{figt2}. 

Since there are no collinear particles in the $\overline{n}^{\mu}$
direction in the final state, we can define the jet function $J_P^a
(\omega, n\cdot k)$ as 
\begin{eqnarray} \label{ja} 
&& \langle 0| T\Biggl[ \frac{1}{\overline{n} \cdot \mathcal{P}}
 \Bigl[ W_n^{\dagger} i\FMSlash{D}_n^{\perp} W_n \Bigr] \Bigl[ \delta
 (\omega -\overline{n} \cdot \mathcal{P}) W_n^{\dagger} \xi_n \Bigr]
 (z) \cdot \Bigl[\overline{\xi}_n W_n \Bigr] 
 \Bigl[ W_n^{\dagger} i\overleftarrow{\FMSlash{D}_n^{\perp}} W_n \Bigr]
 \frac{1}{\overline{n} \cdot \mathcal{P}^{\dagger}} (0) \Biggr] |0\rangle
 \nonumber \\
&& \equiv i\frac{\FMslash{n}}{2} \int \frac{d^4 k}{(2\pi)^4} e^{-ik\cdot
   z} J_P^a (\omega, n\cdot k),
\end{eqnarray}
and $\hat{T}_2^{aa}$ can be written, with $z=\overline{n} \cdot z/2$,
as 
\begin{eqnarray}
\hat{T}_2^{aa} (x,Q) &=& - \int d\omega d\w  B^2_a (\omega, \w) 
\int \frac{dn\cdot k d z}{2\pi} e^{i\Bigl( (1-w) \w/2 -n\cdot 
  k\Bigr) z} J_P^a (\omega, n\cdot k)  \\
&\times& T\Bigl[ \overline{\xi}_{\bar{n}} W_{\bar{n}}
\tilde{S}_{\bar{n}}^{\dagger} S_n (z)
\frac{\FMslash{n}}{2} S_n^{\dagger} S_{\bar{n}} 
\delta (\w - \mathcal{P}_+) W_{\bar{n}}^{\dagger}
\xi_{\bar{n}} (0)\Bigr] \nonumber \\
&=& - \int d\omega d\overline{\omega} B_a^2(\omega, \w)
\int \frac{dn\cdot k d z}{2\pi} \int d\eta e^{i\Bigl( (1-w) \w/2
  -n\cdot 
  k-\eta \Bigr) z} J_P^a (\omega, n\cdot k) \nonumber
\\
&\times& \frac{1}{N} \langle 0| \mathrm{tr} \Bigl[
\tilde{S}_{\bar{n}}^{\dagger} S_n \delta (\eta +n\cdot i\partial)
S_n^{\dagger} S_{\bar{n}} \Bigr] |0\rangle
\cdot T\Bigl[ \overline{\xi}_{\bar{n}} W_{\bar{n}} \delta
(\overline{\omega} -n\cdot \mathcal{P}_+) \frac{\FMslash{n}}{2}
W_{\bar{n}}^{\dagger} \xi_{\bar{n}}\Bigr],
\end{eqnarray}
By putting $\eta = (1-v)n\cdot p$ ($w<v<1$), the spin-averaged matrix
element between the proton state, $T_2^{aa}$ is given as
\begin{eqnarray}
T_2^{aa} (x,Q)&=& - n\cdot P \int d\omega d\overline{\omega}
B_a^2(\omega, \w)
\int d\eta S(\eta) J_P^a \Bigl(\omega, (1-w) \frac{\w}{2} -\eta\Bigr)
\nonumber \\ 
&\times& \int_0^1 dy \delta (\overline{\omega} - 2yn\cdot P) g_P^q (y)
\nonumber \\
&=& -\int_x^1 \frac{dy}{y} g_P^q (y) \int_w^1
dv \tilde{S} (v) \int d\omega B_a^2 (Q,\omega)
n\cdot p J_P^a \Bigl(\omega, (v-w) n\cdot p\Bigr),
\end{eqnarray}

We can also compute the contribution to $T_2(x,Q)$ using
$J_{\mu}^{(1b)}$, which comes from the part proportional to $n_{\mu}
n_{\nu}$ in $\hat{T}_{\mu \nu}$. It is written as
\begin{eqnarray}
\hat{T}_2^{bb} (x,Q) &=& i\int d^4 z e^{i(q-\tilde{p}' +\tilde{p})\cdot z} 
\int d\omega d\omega^{\prime} d\w 
B_b(\omega,\w) B_b(\omega^{\prime},\w)
  \nonumber \\
&\times& T\Biggl[ \overline{\xi}_{\bar{n}} W_{\bar{n}} \frac{1}{n\cdot
  \mathcal{P}^{\dagger}} \Bigl[ W_n^{\dagger}
i\overleftarrow{\FMSlash{D}_n^{\perp}} W_n \Bigr] \Bigl[ \delta
(\omega -\overline{n}\cdot \mathcal{P}) W_n^{\dagger} \xi_n \Bigr] (z)
\nonumber \\
&\times& \Bigl[ \overline{\xi}_n W_n \delta (\omega^{\prime} -
\overline{n} \cdot \mathcal{P}^{\prime}) \Bigr] \Bigl[ W_n^{\dagger}
i\FMSlash{D}_n^{\perp} W_n \Bigr] 
\delta (\w-\mathcal{P}_+) \frac{1}{n\cdot \mathcal{P}}
W_{\bar{n}}^{\dagger} \xi_{\bar{n}} (0)\Biggr].
\end{eqnarray}
By defining the jet function $J_P^b (\omega, n\cdot k)$ as 
\begin{eqnarray} \label{jb}
  && \langle 0| T\Biggl[ \Bigl[W_n^{\dagger}
  i\overleftarrow{\FMSlash{D}_n^{\perp}} W_n\Bigr] \Bigl[ \delta
  (\omega -\overline{n} \cdot \mathcal{P}) W_n^{\dagger} \xi_n \Bigr]
  (z) \cdot \Bigl[ \overline{\xi}_n W_n\Bigr] \Bigl[ W_n^{\dagger} 
i\FMSlash{D}_n^{\perp} W_n \Bigr] (0) \Biggr]|0\rangle
  \nonumber \\
&&\equiv i\frac{\FMslash{n}}{2} \int \frac{d^4 k}{(2\pi)^4}
e^{-ik\cdot z} J_P^b (\omega, n\cdot k), 
\end{eqnarray}
$\hat{T}_2^{bb}$ after factorizing the usoft interactions is
given as 
\begin{eqnarray}
\hat{T}_2^{bb} (x,Q) &=& -\int d\omega d\overline{\omega} B_b^2 (\omega,\w)
\int  \frac{dn\cdot k d z}{2\pi} e^{i\Bigl( (1-w)
  \w/2 -n\cdot k\Bigr) z} J_P^b (\omega, n\cdot k) \\
&\times& T\Bigl[ \overline{\xi}_{\bar{n}} W_{\bar{n}}
\tilde{S}_{\bar{n}}^{\dagger} S_n (z) \frac{\FMslash{n}}{2}
S_n^{\dagger} S_{\bar{n}} \delta 
(\overline{\omega} -n\cdot \mathcal{P}_+) W_{\bar{n}}^{\dagger}
\xi_{\bar{n}} (0) \Bigr] \nonumber \\
&=& -\int d\omega d\overline{\omega} B_b^2 (\omega,\w) \int
 \frac{dn\cdot k d z}{2\pi} \int d\eta e^{i\Bigl(
   (1-w) \w/2 -n\cdot k -\eta \Bigr) z} J_P^b
 (\omega, n\cdot k) \nonumber \\
&\times& \frac{1}{N} \langle 0| \mathrm{tr} \Bigl(
\tilde{S}_{\bar{n}}^{\dagger} S_n \delta (\eta +n\cdot i\partial)
S_n^{\dagger} S_{\bar{n}} \Bigr) |0\rangle \cdot  
\overline{\xi}_{\bar{n}} W_{\bar{n}} \frac{1}{n\cdot
  \mathcal{P}^{\dagger}} \frac{\FMslash{n}}{2} \delta
(\overline{\omega} -n\cdot \mathcal{P}_+) \frac{1}{n\cdot \mathcal{P}}
W_{\bar{n}}^{\dagger} \xi_{\bar{n}}, \nonumber
\end{eqnarray}
and the spin-averaged matrix element between the proton state,
$T_2^{bb}$, is given as 
\begin{eqnarray}
T_2^{bb} (x,Q)&=& -n\cdot P \int d\omega d\w B_b^2 (\omega,\w) \int
d\eta S(\eta) J_P^b \Bigl(\omega, (1-w)\w/2 -\eta\Bigr) \nonumber \\
&\times& \frac{4}{\w^2} \int_0^1 dy \delta (\overline{\omega} -
2yn\cdot P)   g_P^q (y) \nonumber \\
&=& - \int_x^1 \frac{dy}{y} g_P^q (y) \int_w^1 dv
\tilde{S} (v) \int d\omega \frac{B_b^2 (Q,\omega)}{n\cdot p} J_P^b
\Bigl(\omega, (v-w) n\cdot p\Bigr).  
\end{eqnarray}
We can clearly see that both $T_2^{aa}$ and $T_2^{bb}$ factorize. 
By taking the imaginary part of $T_2^{aa}$ or $T_2^{bb}$, the
longitudinal structure function $F_L (x,Q)$ can be written as
\begin{eqnarray} \label{flf}
 F_L (x,Q) &=& 8  \int_x^1 \frac{dy}{y} g_P^q (y)
 \int_w^1 dv \tilde{S} (v) \int d\omega B_a^2 (Q,\omega) 
n\cdot p \Bigl[\frac{-1}{\pi} 
 \mathrm{Im} J_P^a \Bigl( \omega, (v-w) n\cdot p \Bigr)\Bigr],
 \nonumber \\
 F_L (x,Q) &=& 8\int_x^1 \frac{dy}{y} g_P^q (y) \int_x^1 dv \tilde{S}
 (v) \int d\omega \frac{B_b^2 (Q,\omega)}{n\cdot p}
 \Bigl[\frac{-1}{\pi} \mathrm{Im} J_P^b \Bigl( \omega, (v-w) n\cdot
 p\Bigr) \Bigr],   
\end{eqnarray}
where the first (second) expression is obtained using $T_2^{aa}$
($T_2^{bb}$).

\begin{figure}[t]
\begin{center}
\epsfig{file=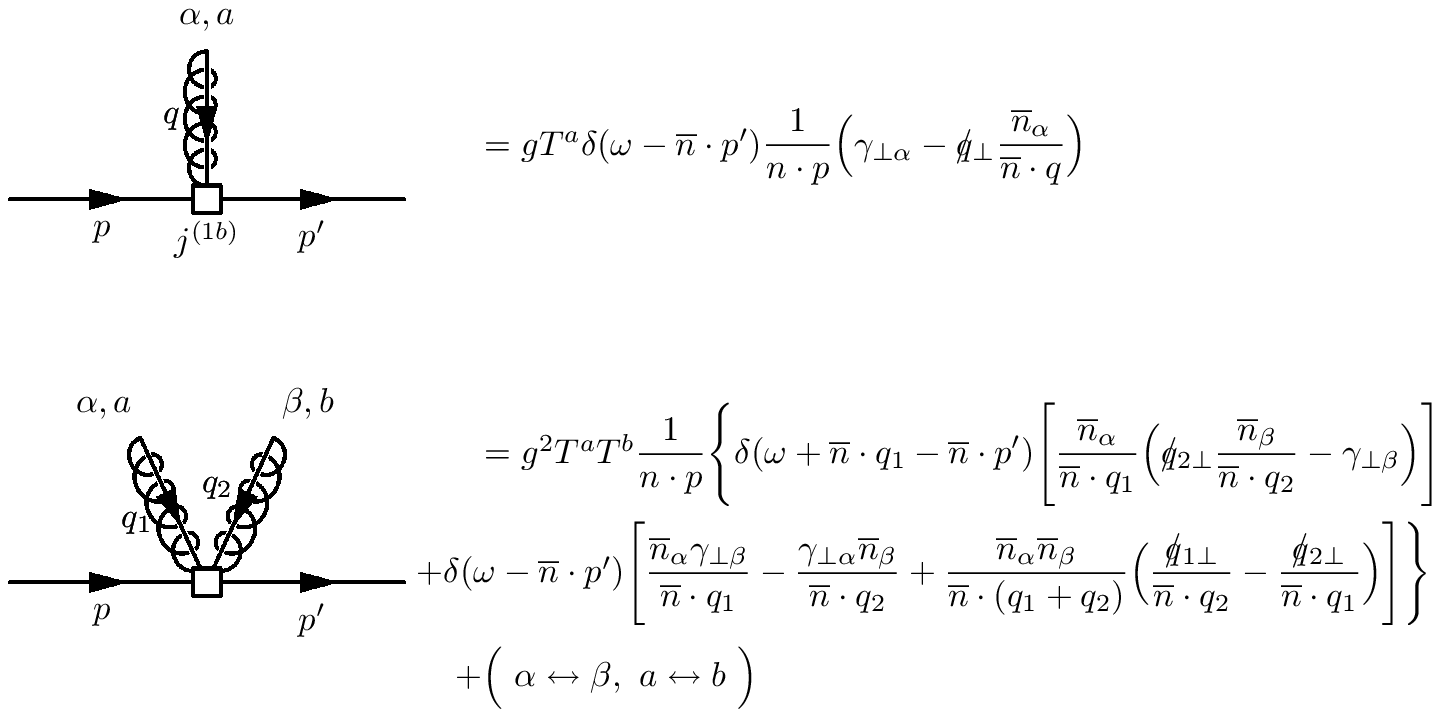, width=13cm}
\end{center}
\vspace{-1.0cm}
\caption{Feynman rules of the current $j^{(1b)}(\omega)$ with one or two
$n$-collinear gluons.}
\label{figjb}
\end{figure}

\begin{figure}[b]
\begin{center}
\epsfig{file=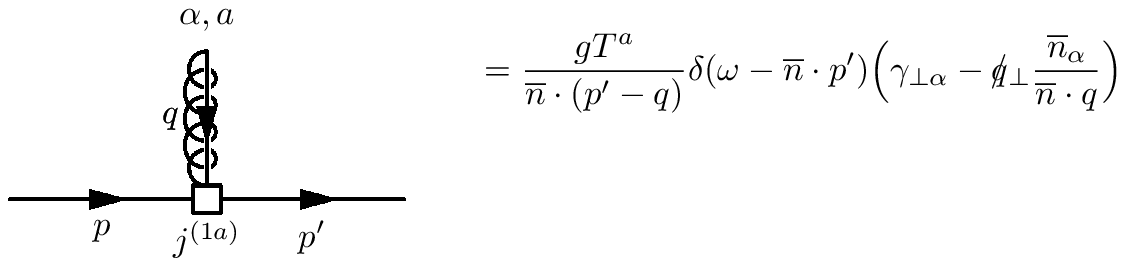, width=10.5cm}
\end{center}
\vspace{-1.0cm}
\caption{Feynman rules of the current $j^{(1a)}(\omega)$ with a single
$n$-collinear gluon.}
\label{figjar}
\end{figure}

The jet function $J_P^b (\omega, n\cdot k)$ at order $\alpha_s$ can be
computed from Eq.~(\ref{jb}). The Feynman rules for $j^{(1b)}$ with
one or two collinear gluons in the $n^{\mu}$ direction are shown in
Fig.~\ref{figjb}. And the matrix element $M_b$ from Fig.~\ref{figt2}
with $J_{\mu}^{(1b)}$, after extracting the overall factor $1/(n\cdot
p)^2$, is given as  
\begin{eqnarray}
  M_b &=& g^2 C_F \int \frac{d^D l}{(2\pi)^D}
  \gamma_{\perp\alpha} 
  \delta \Bigl(\omega -\overline{n}\cdot (l+p^{\prime}) \Bigr)
  \frac{\FMslash{n}}{2} \frac{\overline{n} \cdot
    (l+p^{\prime})}{(l+p^{\prime})^2} \gamma_{\perp}^{\alpha}
  \frac{1}{l^2} \nonumber \\
&=& -g^2 C_F \frac{\FMslash{n}}{2} (D-2) \int  \frac{d^D l}{(2\pi)^D}
\delta \Bigl(\omega -\overline{n}\cdot (l+p^{\prime}) \Bigr)
\frac{\overline{n} \cdot     (l+p^{\prime})}{l^2 (l+p^{\prime})^2}
\nonumber \\
&\longrightarrow& i\frac{\FMslash{n}}{2} \frac{\alpha_s C_F}{2\pi}
\frac{\omega}{\overline{n} \cdot p^{\prime}} \theta (\omega) \theta
  (\overline{n} \cdot p^{\prime} -\omega)  \Bigl[ 1+\ln \frac{\omega
    (\overline{n} \cdot p^{\prime} -\omega) (-n\cdot k
    -i\epsilon)}{\overline{n} \cdot p^{\prime} \mu^2} \Bigr], 
\end{eqnarray}
where we collect the finite terms only in the last
expression. Here $\overline{n} \cdot p^{\prime}$ is the total collinear
momentum of the intermediate states, and can be replaced by
$\overline{n}\cdot q =Q$. Therefore the jet function $J_P^b (\omega,
n\cdot k)$ is given by
\begin{equation}
J_P^b (\omega, n\cdot k) =  \frac{\alpha_s C_F}{2\pi}
\frac{\omega}{Q} \theta (\omega) \theta
  (Q -\omega)  \Bigl[ 1+\ln \frac{\omega
    (Q -\omega) (-n\cdot k
    -i\epsilon)}{Q \mu^2} \Bigr].  
\end{equation}
Comparing the definitions of $J_P^a$ and $J_P^b$ in
Eqs.~(\ref{ja}) and (\ref{jb}), $J_P^a (\omega, n\cdot k)$ is given by
\begin{equation} \label{jab}
J_P^a (\omega, n\cdot k) =  \frac{1}{Q^2} J_P^b  (\omega, n\cdot k). 
\end{equation}
It is easy to see this relation by looking at the Feynman rules for
$j^{(1a)}$, which are presented in Fig.~\ref{figjar}. Note that there is
an additional factor of $1/(\overline{n}\cdot p^{\prime})^2$ in the
definition of $J_P^a$, and the nonzero contribution comes from the part
proportional to $\gamma_{\perp\alpha}$, of which the radiative
corrections are the same using either $j^{(1a)}$ or $j^{(1b)}$.

Let us simplify the expression for $F_L (x,Q)$ in Eq.~(\ref{flf}). At
order $\alpha_s$, since the jet functions are already at order
$\alpha_s$, the Wilson coefficients $C(Q)$ and $B_{a,b}(\omega)$ take their
tree-level values, that is, 1. Introducing the dimensionless variable
$r=\omega/Q$, the jet function $J_P^b$ is
written as
\begin{equation}
  J_P^b \Bigl( (v-w) n\cdot p\Bigr) =\frac{\alpha_s C_F}{2\pi}
  r\theta(r) \theta (1-r) \Bigl\{ 1 +\ln \Bigr[ \frac{Q^2}{\mu^2} r(1-r)
  \Bigl( -(1-\frac{v}{w}) -i\epsilon\Bigr) \Bigr] \Bigr\}. 
\end{equation}
Therefore the imaginary part is given by
\begin{equation}
  \overline{J}_P^b \Bigl(r, \frac{w}{v} \Bigr) \equiv -\frac{1}{\pi}
  \mathrm{Im} J_P^b \Bigl( (v-w) n\cdot p\Bigr) =\frac{\alpha_s
    C_F}{2\pi} r \theta (r) \theta (1-r). 
\end{equation}
At order $\alpha_s$, the longitudinal structure function $F_L (x,Q)$
becomes
\begin{equation}
  F_L (x,Q) = 8 \frac{\alpha_s C_F}{4\pi} \int_x^1
  \frac{dy}{y} g_P^q (y) \frac{1}{w} \int_w^1 dv   \tilde{S} (v)
= 8  \frac{\alpha_s C_F}{4\pi} \int_x^1  \frac{dy}{y} g_P^q (y)
\int_w^1 dv  w \tilde{S} (v),
\end{equation}
where the first (second) expression comes from $T_2^{aa}$
($T_2^{bb}$). The two expressions differ by a factor of $w^2$, which
gives a subleading correction since $1-w \sim \Lambda/Q$. Therefore 
the contributions from $T_2^{aa}$ and $T_2^{bb}$
to $F_L (x,Q)$ are the same near the endpoint region at leading order
in $\Lambda/Q$. In fact, the result of the electromagnetic current
conservation goes further than the fact that $T_2$ can be obtained
using either $J_{\mu}^{(1a)}$ and $J_{\mu}^{(1b)}$. It should hold to
all orders in $\alpha_s$, and the scaling behavior of the two currents
should also be the same. We present the anomalous dimensions for
$J_{\mu}^{(1b)}$ at one loop in Appendix C. 

The moment of $F_L (x,Q)$ using the first expression in
Eq.~(\ref{flf}) is written as
\begin{eqnarray}
  F_{L,n} (Q) &=& \int_0^1 dx x^{n-1} F_L (x,Q) \nonumber \\
&=& \frac{2\alpha_s C_F}{\pi}\int_0^1 dx x^{n-1} \int_0^1 dy g_P^q
(y) \int_0^1 dw \delta (x-wy) \frac{1}{w} \int_w^1 dv  \tilde{S} (v)
\nonumber \\ 
&=& \frac{2\alpha_s C_F}{\pi}  \int_0^1 dy y^{n-1} g_P^q (y)
\int_0^1 dw w^{n-2} \int_w^1 \frac{dv}{v} v\tilde{S} (v) \nonumber \\
&=& \frac{2\alpha_s C_F}{\pi}  \int_0^1 dy y^{n-1} g_P^q (y)
\int_0^1 dw w^{n-2} \int_0^1 du \int_0^1 dv \delta (w-vu) v\tilde{S}
(v)  \nonumber \\
&=&  \frac{2\alpha_s C_F}{\pi} \frac{1}{n-1} \int_0^1 dy y^{n-1}
g_P^q (y) \int_0^1 dv v^{n-1} \tilde{S} (v)  \nonumber \\
&\longrightarrow& \frac{2\alpha_s C_F}{\pi} \frac{1}{n} g_{P,n}^q 
\cdot \tilde{S}_n = \frac{2\alpha_s C_F}{\pi} \frac{1}{n}
f_{P,n,n+1}^q,  
\end{eqnarray}
where the last expression is obtained in the large $n$ limit. Compared
to the moment $F_{1,n}$, $F_{L,n}$ is suppressed by $n$, which confirms
that the longitudinal structure function $F_L (x,Q)$ is suppressed by
$\Lambda/Q$ compared to $F_1 (x,Q)$. 
In order to obtain the exponentiated form, we should compute the
radiative corrections to next-to-leading order accuracy. This has not
been done here, but all the logarithmic terms 
such as $\alpha_s^k \ln^l n /n$ can be resummed from  
the factorization property in SCET, as suggested in the approach using
the full theory \cite{Akhoury:1998gs}.

\section{Conclusion\label{concl}}
DIS near the endpoint region can be described in
SCET. The factorization of the structure functions is explicitly shown
to order $\alpha_s$, and the moments of the structure functions are
expressed as a product of the Wilson coefficients, the moments of the
jet functions, the collinear matrix elements, and the soft Wilson
lines. The radiative corrections for each component can be separately
computed using perturbation theory. The structure function $F_1 (x,Q)$
starts at leading order in $\Lambda$, while the longitudinal structure
function $F_L (x,Q)$ starts from order $\Lambda$ and
$\alpha_s$. Therefore the Callan-Gross sum rule holds at leading order
in $\Lambda$, and the corrections can be systematically computed in
SCET. 

High-energy processes, such as DIS, Drell-Yan
processes, hadron collisions, $e^+e^- \rightarrow $ jets, can be
described by SCET. And all these processes possess common
features though the detailed dynamics are different. First, the
scattering cross sections are factorized, and the
short-distance physics and the long-distance physics are
separated. There is a universal soft Wilson line describing soft 
gluon emissions near the endpoint region with the appropriate
prescription for the soft Wilson lines depending on the external
collinear particles. And there is a contribution from the  parton
distribution functions if the initial particles are hadrons. 

Compared to the approach in full QCD \cite{css}, there is advantage in
employing SCET in DIS near the endpoint
region. First, the factorization property becomes transparent since
SCET is formulated from the beginning to decouple the collinear and
the soft degrees of freedom. And the power counting in powers of
$\Lambda$ can be systematically performed. Secondly, once the
factorized form is given, each factor has a different physical origin
and its radiative corrections and evolutions can be computed in
perturbation theory. The hard coefficient $C(Q,\mu_0)$ comes from the
hard physics of order $Q$ and can be computed in matching the full
theory and $\mathrm{SCET}_{\mathrm{I}}$. The jet function $J_P$ arises
from the hard-collinear physics of order $\sqrt{Q\Lambda}$, and can be
computed by matching $\mathrm{SCET}_{\mathrm{I}}$ and
$\mathrm{SCET}_{\mathrm{II}}$. The soft Wilson lines and the collinear
matrix elements are combined to give the parton distribution
functions.  Since
the collinear particles and the soft particles are decoupled, the
radiative corrections for the soft Wilson line are governed only by the
soft interactions in $\mathrm{SCET}_{\mathrm{II}}$, while those for
the collinear matrix elements are governed only by the collinear
interactions. To guarantee this, and to avoid double counting, the
zero-bin subtraction should be performed. Each contribution is clearly
separated and SCET specifies the prescription for computing radiative
corrections.  

In this paper we have considered the flavor nonsinglet structure
functions. To be complete, the flavor singlet structure functions
should be included. In this case we have to consider the collinear
operators with gluons, which contribute to the gluon distribution
function in the proton. Though it will be more involved because of the
operator mixing, the procedure for showing the factorization is
straightforward. The complete treatment of DIS near the endpoint
region including the flavor singlet structure function will be
considered elsewhere. It will be interesting to see if other various
high-energy processes near the endpoint region can have similar
features as those in DIS.

Note added: The original preprint version of this paper did not
include the zero-bin contribution \cite{Manohar:2006nz}, and the
parton distribution 
function in SCET was not properly defined excluding the soft
part. In this paper, the zero-bin subtraction is correctly performed
to solve the double counting problem, and the parton distribution
function is carefully defined in the endpoint region. Because of
these, the results of the calculations and the conclusion
of the paper have changed. However, in the meantime, several authors
\cite{Becher:2006mr, Chen:2006vd} criticized this paper based on the
original preprint version. We would like to comment on the criticism
which appeared in the literature.

In Ref.~\cite{Becher:2006mr}, the authors argue that there should be
no extra soft contributions outside the parton distribution
function, which is correct based on the original
preprint. However, in this paper, we include the soft contributions as
part of the parton distribution function, which also affects the jet
function in the endpoint region, contrary to the case away from the
endpoint.  They also claim that the
$\overline{n}$-collinear gluon exchange shown in Fig.~\ref{figs1hc}
(b), (c) are not kinematically allowed. But all the
$\overline{n}$-collinear contributions in Fig.~\ref{figs1hc} should be
included because the interaction of the collinear gluon with the quark
happens inside the proton, and the resulting momentum of the quark
undergoing the hard collision should be regarded as $n$ collinear, not
the momentum of the quark before it interacts with a collinear
gluon. The detail is explained in Appendix A.

The main criticism of Ref.~\cite{Chen:2006vd} is that the double
counting problem is not performed properly, and that the parton
distribution function is not correctly identified if only the matrix
elements of the collinear operators are included. In the current
paper, the double counting is treated using the zero-bin subtraction
method. For the parton distribution function, we include the effect of
the soft gluons in the parton distribution function. Besides these,
we agree with their claim that the structure function depends
only on $Q^2$, and $(1-x)Q^2$, not on $(1-x)^2 Q^2$, which is
illustrated clearly in Eq.~(\ref{f1final}), where the structure
function depends  only on $Q^2$ and 
$\mu_0^2 = (1-x)Q^2$, and the dependence on $\mu$ cancels.  

\section*{Acknowledgments}
J.~C. was supported by the Korea Research Foundation Grant
KRF-2005-015-C00103. C.~K. was supported by the National Science
Foundation under Grant No. PHY-0244599. 

\appendix

\section{Zero-bin subtraction before the usoft factorization}
For the purpose of illustrating the zero-bin subtraction
method in $\mathrm{SCET}_{\mathrm{I}}$, we consider the radiative
corrections to the forward 
scattering amplitude before the usoft factorization. The starting
point is that $\hat{T}_{\mu\nu}^{(0)}$ from Eq.~(\ref{tmunu0}) is
given by
\begin{eqnarray} 
\hat{T}^{(0)}_{\mu\nu} &=& i   C^2(Q) \int d^4 z
  e^{i(q+\tilde{p}-\tilde{p}^{\prime}) 
  \cdot z} T\Bigl[ J_{\mu}^{(0)\dagger} (z) J_{\nu}^{(0)} (0)\Bigr]
\nonumber \\
&=& i   C^2(Q) \int d^4 z
  e^{i(q+\tilde{p}-\tilde{p}^{\prime}) 
  \cdot z} T\Bigl[ \overline{\xi}_{\bar{n}} W_{\bar{n}}
 \gamma_{\mu} W_n^{\dagger} \xi_n (z) \overline{\xi}_n W_n
 \gamma_{\nu}   W_{\bar{n}}^{\dagger} \xi_{\bar{n}} (0) \Bigr], 
\end{eqnarray}
where the collinear field is not redefined by Eq.~(\ref{usoftd}), and
can interact with usoft gluons.

\begin{figure}[t]
\begin{center}
\epsfig{file=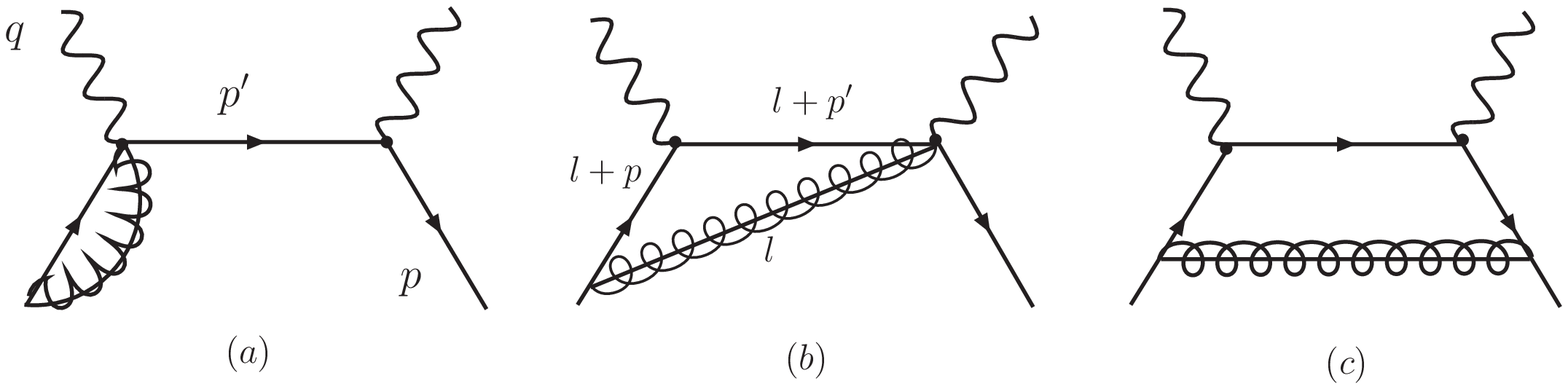, width=13cm}
\end{center}
\vspace{-1.0cm}
\caption{Feynman diagrams of the contribution from collinear gluons to
  the forward scattering amplitude in $\rm{SCET_I}$. The mirror images
  of (a) and (b) are omitted.} 
\label{figs1hc}
\end{figure}

We consider the radiative corrections with the
$\overline{n}$-collinear gluons, which contribute to the
renormalization of the quark distribution functions. The Feynman
diagrams are shown in Fig.~\ref{figs1hc}. We consider the amplitudes
only at the parton level,and the convolution with the parton
distribution function is straightforward. The naive contributions
proportional to $-g^{\perp}_{\mu\nu}$ from Fig.~\ref{figs1hc} (a), (b)
with their mirror images and (c) are given as  
\begin{eqnarray} \label{colc}
  M_a &=&4ig^2 C_F   \frac{\FMslash{n}}{2} 
  \frac{1}{n\cdot p^{\prime}} \int \frac{d^D l}{(2\pi)^D}
  \frac{n\cdot  (l+p)}{l^2 (l+p)^2 n\cdot l}, \nonumber \\ 
M_b &=& -4ig^2 C_F \frac{\FMslash{n}}{2} \int \frac{d^D l}{(2\pi)^D}
\frac{n\cdot (l+p)}{l^2   (l+p)^2 n\cdot (l+p^{\prime}) n\cdot l},
\nonumber \\  
M_c &=& ig^2 C_F \frac{\FMslash{n}}{2} \int \frac{d^D l}{(2\pi)^D}
\frac{1}{l^2   [(l+p)^2]^2 n\cdot (l+p^{\prime})} \gamma_{\perp
  \alpha} \FMslash{l}_{\perp}  \FMslash{l}_{\perp}
\gamma_{\perp}^{\alpha},  
\end{eqnarray}
where $p^{\prime}_{\mu} = p_{\mu}+q_{\mu}$ is the $n$-collinear
momentum, and we put $p_{\perp}=0$ for simplicity.

\begin{figure}[b]
\begin{center}
\epsfig{file=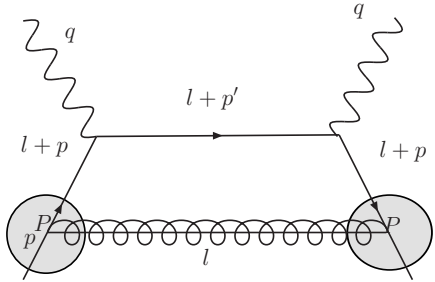, width=6cm}
\end{center}
\vspace{-1.0cm}
\caption{When an $\overline{n}$ gluon is exchanged, there is a
  kinematical region where the initial state with $l+p$ is
  $\overline{n}$ collinear, and the intermediate state with
  $l+p^{\prime}$ is  $n$ collinear, while the loop momentum $l$ is
  $\overline{n}$ collinear.} 
\label{kinematics1}
\end{figure}

Note that in $M_b$, the propagators are written 
in such a way that $l+p^{\prime}$ is collinear in the $n$
direction. This should be included in the collinear contribution,
contrary to the claim in Ref.~\cite{Becher:2006mr}, in which the
authors claim that 
Fig.~\ref{figs1hc} (b), and (c) should not be included since they are
kinematically forbidden. However, by looking into the kinematics
carefully, there are collinear contributions from Fig.~\ref{figs1hc}
(b), and (c). The point is that a quark inside the proton can interact
with $\overline{n}$-collinear gluons before the hard collision. 
Therefore the $\overline{n}$-collinear gluon is regarded as part 
of the proton, and forms an $\overline{n}$ collinear jet. The parton
distribution function describes the partons which
undergo a hard collision after all the interactions with the collinear
jet. 

The situation is schematically shown in Fig.~\ref{kinematics1}, which
is Fig.~\ref{figs1hc} (c). The
collinear quark interacts with a collinear gluon before it collides
with a hard photon. Therefore the longitudinal momentum of the
collinear quark for the hard collision is $n\cdot (l+p)$, not $n\cdot
p$, where $n\cdot p$ is the longitudinal momentum fraction before it
interacts with a collinear gluon. In order to see if the
$\overline{n}$-collinear gluon is allowed by kinematics, let us
introduce the partonic variable $w^{\prime}$, to avoid confusion,
which is given by
\begin{equation}
  w^{\prime} \sim -\frac{n\cdot q}{n\cdot (l+p)}.
\end{equation}
And $n\cdot (l+p)$ and $n\cdot (l+p^{\prime})$ are given by 
\begin{equation}
n\cdot (l+p) = Q/w^{\prime}, \ \  n\cdot (l+p^{\prime}) = n\cdot
(l+p+q) = (1-w^{\prime}) n\cdot (l+p) =
\frac{1-w^{\prime}}{w^{\prime}} Q.   
\end{equation}
In the endpoint region where $w^{\prime} \rightarrow 1$, 
$l+p$ can be  $\overline{n}$ collinear, $l+p^{\prime}$ can be $n$
collinear, while $l$ is $\overline{n}$ collinear. Therefore the
contribution from $\overline{n}$-collinear gluons should be included.

The reason why Eq.~(\ref{colc}) is naive is because the loop momentum
can be soft, which should be avoided in the collinear
sector. Therefore we subtract the contribution where the loop 
momentum becomes soft, and we call this the zero-bin contribution. 
It can be obtained from Eq.~(\ref{colc}) by power counting, where  all
the components of the loop momentum $l^{\mu}$ scales as $\Lambda$. The
zero-bin amplitudes are given as
\begin{eqnarray}
M_a^0 &=& 4ig^2 C_F \frac{\FMslash{n}}{2} \frac{1}{n\cdot p^{\prime}}
\int \frac{d^D l}{(2\pi)^D} 
\frac{n\cdot  p}{l^2 (n\cdot p \overline{n} \cdot l +p^2) n\cdot l},
\nonumber    \\ 
M_b^0 &=& -4ig^2 C_F \frac{\FMslash{n}}{2} \int \frac{d^D l}{(2\pi)^D}
\frac{n\cdot p}{l^2 (n\cdot p   \overline{n}\cdot l + p^2) n\cdot
  (l+p^{\prime}) n\cdot l}, 
\end{eqnarray}
where we put $p^2$ to regulate the infrared divergence. And the
zero-bin contribution from Fig.~\ref{figs1hc} (c) is suppressed, and
we neglect it here. The total zero-bin contribution is given as
\begin{eqnarray}
  M_a^0 + M_b^0 &=& 4ig^2 C_F \frac{\FMslash{n}}{2}  \int \frac{d^D
    l}{(2\pi)^D} \frac{n\cdot p}{l^2 (n\cdot p \overline{n} \cdot l
    +p^2) n\cdot l} \Bigl[   \frac{1}{n\cdot     p^{\prime}} -
  \frac{1}{n\cdot (l+p^{\prime})}   \Bigr] \nonumber  \\
&=&  4ig^2 C_F \frac{\FMslash{n}}{2} \frac{1}{n\cdot p^{\prime}} \int
\frac{d^D  l}{(2\pi)^D} \frac{n\cdot p}{l^2  (n\cdot p
  \overline{n}\cdot l + p^2) n\cdot (l+p^{\prime})}. 
\end{eqnarray}

\begin{figure}[b]
\begin{center}
\epsfig{file=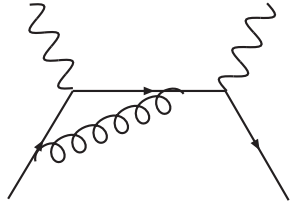, width=4cm}
\end{center}
\vspace{-1.0cm}
\caption{Feynman diagram of the contribution from usoft gluons to
  the forward scattering amplitude in $\rm{SCET_I}$, and the mirror
  image is omitted.} 
\label{figs1us}
\end{figure}

The usoft contributions from Fig.~\ref{figs1us} are given by
\begin{equation}
M_{us} =  4ig^2 C_F \frac{\FMslash{n}}{2} \frac{1}{n\cdot p^{\prime}}
\int \frac{d^D l}{(2\pi)^D} \frac{n\cdot p}{l^2  (n\cdot p
  \overline{n}\cdot l + p^2) n\cdot  (l+p^{\prime})}, 
\end{equation}
which is exactly equal to the zero-bin contribution. Therefore the
correct computation including the zero-bin subtraction becomes
\begin{equation}
M_a +M_b +M_c -(M_a^0 +M_b^0 )  +M_{us}  =M_a +M_b+M_c,  
\end{equation}
which states that the naive collinear contribution without the
zero-bin subtraction gives the correct result. It is also true in
$\mathrm{SCET}_{\mathrm{II}}$ after the soft factorization. The
zero-bin contribution to the collinear operator is the same as the
radiative correction to the soft Wilson line. 

In calculating $M_a +M_b$, note that we can write $M_b$ as 
\begin{eqnarray}
 M_b &=& -4ig^2 C_F \frac{\FMslash{n}}{2} \int \frac{d^D l}{(2\pi)^D}
 \frac{n\cdot (l+p)}{l^2 (l+p)^2 n\cdot   (l+p^{\prime}) n\cdot l}
 \nonumber \\ 
&=& - 4ig^2 C_F \frac{\FMslash{n}}{2} 
 \frac{1}{n\cdot p^{\prime}} \int \frac{d^D l}{(2\pi)^D} \Bigl[
 \frac{n\cdot (l+p)}{l^2 (l+p)^2 n \cdot l} -\frac{n\cdot (l+p)}{l^2
   (l+p)^2 n\cdot (l+p^{\prime})}\Bigr], 
\end{eqnarray}
where the first term is equal to $-M_a$. Therefore $M_a +M_b$ is given
as 
\begin{equation}
M_a +M_b = 4ig^2 C_F \frac{\FMslash{n}}{2} 
\frac{1}{n\cdot p^{\prime}} \int \frac{d^D l}{(2\pi)^D} \frac{n\cdot
(l+p)}{l^2    (l+p)^2 n\cdot (l+p^{\prime})}.  
\end{equation}
Evaluating the $\overline{n} \cdot l$ integral by contours, doing the
${\bf l}_{\perp}$ integral, and using the substitution $n\cdot l= -z
n\cdot p$ gives the infinite part
\begin{eqnarray}
M_a +M_b &=&- \frac{\alpha_s C_F}{\pi} \frac{1}{\epsilon}
\frac{\FMslash{n}}{2} \frac{1}{n\cdot p}  \int_0^1 dz
\frac{1-z}{(1-w+i0^+)(1-z-w +i0^+)} 
\nonumber \\ 
&=&  -\frac{\alpha_s C_F}{\pi} \frac{1}{\epsilon}
\frac{\FMslash{n}}{2} \frac{1}{n\cdot p} \frac{1}{1-w  +i0^+} \Bigl( 1
+w \ln \frac{1-w +i0^+}{-w +i0^+}\Bigr),     
\end{eqnarray}
where $n\cdot p^{\prime}=(1-w) n\cdot p$. Similarly, $M_c$ is given as 
\begin{eqnarray}
M_c &=& - \frac{\alpha_s C_F}{2 \pi} \frac{1}{\epsilon}
\frac{\FMslash{n}}{2} \frac{1}{n\cdot p} \int_0^1 dz \frac{z}{1-w -z
  +i0^+} \nonumber \\
&=&  - \frac{\alpha_s C_F}{2 \pi} \frac{1}{\epsilon} 
\frac{\FMslash{n}}{2} \frac{1}{n\cdot p} \Bigl( 1+ (1-w)
\ln \frac{1-w+i0^+}{-w+i0^+} \Bigr).   
\end{eqnarray}
And the wave function renormalization for the external quarks is given
as
\begin{equation}
M_{\mathrm{w.f.}} = \frac{\alpha_s C_F}{4\pi} \frac{1}{\epsilon}
\frac{1}{n\cdot p} \frac{\FMslash{n}}{2} \frac{1}{1-w+i0^+}.   
\end{equation}

Noting that 
\begin{equation}
\mathrm{Im} \, \frac{w}{1-w+i0^+} \ln (w-1-i0^+) = -\pi
\frac{w}{(1-w)_+}, \  \mathrm{Im} \, \frac{1}{1-w +i0^+} = -\pi \delta
(1-w),  
\end{equation}
we obtain
\begin{equation} \label{sc11}
\frac{1}{\pi} \mathrm{Im} \, (M_a +M_b+ M_c +M_{\mathrm{w.f.}})
=\frac{\alpha_s C_F}{2\pi} \frac{1}{\epsilon} 
\frac{1}{n\cdot p} \frac{\FMslash{n}}{2} 
\Bigl( \frac{3}{2} \delta (1-w) +\frac{1+w^2}{(1-w)_+}\Bigr). 
\end{equation}
The tree-level amplitude is given by
\begin{equation}
M_{\mathrm{tree}} = -\frac{1}{n\cdot p} \frac{\FMslash{n}}{2}
\frac{1}{1-w+i0^+},   
\end{equation}
the imaginary part of which is 
\begin{equation} \label{sc1tr}
\frac{1}{\pi} \mathrm{Im} \,  M_{\mathrm{tree}} =  \frac{1}{n\cdot p}
\frac{\FMslash{n}}{2} \delta (1-w). 
\end{equation}
Adding Eqs.~(\ref{sc11}) and (\ref{sc1tr}), we have
\begin{equation}
\frac{1}{n\cdot p} \frac{\FMslash{n}}{2} \Bigl[ \delta (1-w) +
\frac{\alpha_s C_F}{2\pi} \frac{1}{\epsilon} \Bigl( \frac{3}{2} \delta
(1-w) +\frac{1+w^2}{(1-w)_+}\Bigr) \Bigr],  
\end{equation}
from which the anomalous dimension is given as
\begin{equation}
\gamma = -\frac{\alpha_s C_F}{\pi} \Bigl( \frac{3}{2} \delta (1-w)
+\frac{1+w^2}{(1-w)_+}\Bigr), 
\end{equation}
which is exactly the Altarelli-Parisi kernel. Note that this is the
result including the zero-bin subtraction, and it corresponds
to the radiative corrections for the sum of the collinear matrix
element and the soft part. 

\section{Discontinuity in $\mathrm{SCET}_{\mathrm{I}}$  
and   $\mathrm{SCET}_{\mathrm{II}}$} 
It is possible to take the imaginary part of the forward scattering
amplitude to obtain the structure function in
$\mathrm{SCET}_{\mathrm{I}}$ as well as in
$\mathrm{SCET}_{\mathrm{II}}$. If we 
only consider the collinear interactions with the intermediate state,
the computation produces the jet function and the discontinuity due to
the collinear interactions is the same both in
$\mathrm{SCET}_{\mathrm{I}}$ and
$\mathrm{SCET}_{\mathrm{II}}$. Therefore the issue here is 
how to take the discontinuity related to the (u)soft
interactions. We consider the (u)soft interactions with the
intermediate state in both effective theories and show that the
discontinuity is the same. Since we are interested in computing the
anomalous dimension of the soft Wilson line, we focus on the
ultraviolet divergent part.  

\begin{figure}[b]
\begin{center}
\epsfig{file=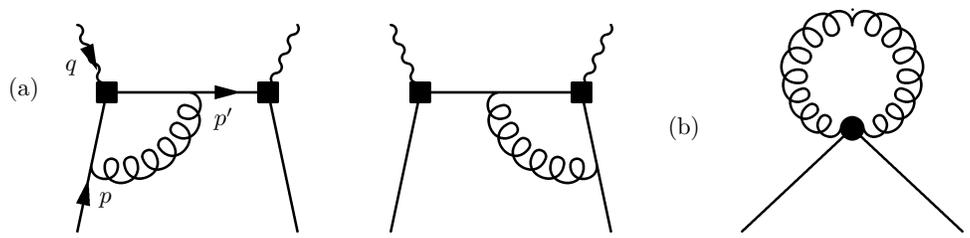, width=13cm}
\end{center}
\vspace{-1.0cm}
\caption{Feynman diagrams for computing the radiative corrections of
  the (u)soft Wilson line in (a) $\mathrm{SCET}_{\mathrm{I}}$ and (b)
  $\mathrm{SCET}_{\mathrm{II}}$.} 
\label{img}
\end{figure}

Let us consider the usoft interactions in $\mathrm{SCET}_{\mathrm{I}}$
and take the discontinuity. The relevant Feynman diagrams are shown in
Fig.~\ref{img} (a), where the curly lines are soft gluons. Using the
dimensional regularization, the Feynman diagrams in Fig.~\ref{img}
(a) are given as 
\begin{equation}
M_a = -4ig^2 C_F O_{\bar{n}}\int \frac{d^D l}{(2\pi)^D}
\frac{1}{(l^2   +i0^+ ) \Bigl(\overline{n}\cdot (l+p) +i 0^+ \Bigr)
  \Bigl(n\cdot  (l+p^{\prime}) +i0^+\Bigr) (n\cdot p^{\prime}+i0^+)},  
\end{equation}
where $O_{\bar{n}}$ is the operator 
\begin{equation}
  O_{\bar{n}} = \overline{\xi}_{\bar{n}} W_{\bar{n}} \gamma_{\mu}
  \frac{\FMslash{n}}{2} \gamma_{\nu} W_{\bar{n}}^{\dagger}
  \xi_{\bar{n}}, 
\end{equation}
Evaluating the $\overline{n} \cdot l$ integral by contours, doing the
${\bf l}_{\perp}$ integral give the infinite part 
\begin{equation}
M_a = \frac{\alpha_s C_F}{\pi \epsilon} \frac{O_{\bar{n}}}{n\cdot
  p^{\prime} +i0^+} \int_{-\infty}^0 dn\cdot l \frac{1}{n\cdot
  (l+p^{\prime} +i0^+)}.   
\end{equation}
By putting $n\cdot p^{\prime} = n\cdot (p+q) = (1-w) n\cdot p$ and
$n\cdot l = -zn\cdot p$, we obtain 
\begin{equation}
M_a = \frac{\alpha_s C_F}{\pi \epsilon}  \frac{O_{\bar{n}}}{n\cdot
    p} \frac{1}{1-w+i0^+} \int_0^1 dz \frac{1}{1-w-z+i0^+}
= \frac{\alpha_s C_F}{\pi \epsilon} \frac{O_{\bar{n}}}{n\cdot
    p} \frac{\ln (w-1-i0^+)}{1-w+i0^+}, 
\end{equation}
where we neglect the $\ln w$ term as $w\rightarrow 1$. The
discontinuity in $\mathrm{SCET}_{\mathrm{I}}$ from the usoft
interactions is given by  
\begin{equation} \label{imma}
  \frac{1}{\pi} \mathrm{Im} \, M_a  = -\frac{\alpha_s
    C_F}{\pi \epsilon} \frac{1}{(1-w)_+} \frac{O_{\bar{n}}}{n\cdot
    p}. 
\end{equation}
This analysis is similar to the analysis in Ref.~\cite{Manohar:2003vb},
in which a single-step matching was performed. 

In $\mathrm{SCET}_{\mathrm{II}}$, the Feynman diagram is shown in
Fig.~\ref{img} (b). It gives
\begin{equation}
  M_b = \int d\eta \frac{1}{(1-w) n\cdot p -\eta +i0^+}
  \frac{\alpha_s C_F}{\pi \epsilon} \frac{\theta(\eta)}{\eta_+}
  O_{\bar{n}},  
\end{equation}
where the first term in the denominator is the coefficient (jet
function at tree level) with the energy transfer $\eta$ to the soft
gluon. The remaining part is the result of the soft loop
calculation ~\cite{Chay:2004zn}. Taking the imaginary part of $M_b$,
we have
\begin{equation}
  \frac{1}{\pi} \mathrm{Im} \, M_b = -\frac{\alpha_s C_F}{\pi \epsilon}
  \frac{1}{(1-w)_+} \frac{O_{\bar{n}}}{n\cdot  p},
\end{equation}
which is the same result as Eq.~(\ref{imma}) obtained in
$\mathrm{SCET}_{\mathrm{I}}$. 

\section{Anomalous dimension of $J_{\mu}^{(1b)}$}
\begin{figure}[b]
\begin{center}
\epsfig{file=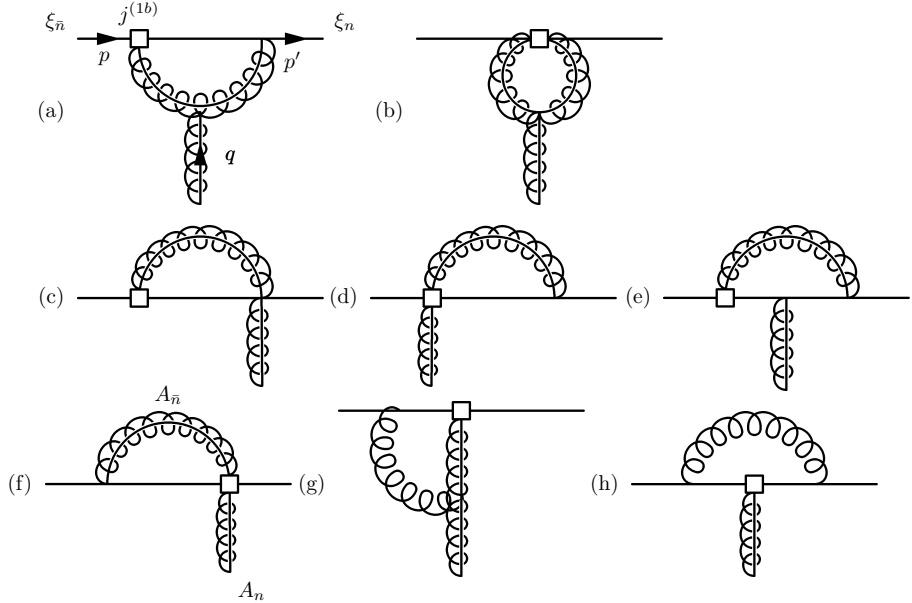, width=12cm}
\end{center}
\vspace{-1.0cm}
\caption{Feynman diagrams for the radiative corrections of  
$j^{(1b)}(\omega)$ in $\rm{SCET_I}$ at one loop. $p$ is in the
$\overline{n}^{\mu}$ direction, and $p^{\prime}$, $q$ are in the $n^{\mu}$
direction ($q$ incoming). Diagrams (a) to (e) include $n$-collinear
loop, (f) includes the $\overline{n}$-collinear loop, and (g), (h) are
the soft corrections.}  
\label{figjbr}
\end{figure}

We present the calculation of the anomalous dimension for
$J_{\mu}^{(1b)}$ at one loop, and explain why it is the same for 
$J_{\mu}^{(1a)}$. The current $J_{\mu}^{(1b)}$ from
Eq.~(\ref{j1ab}) is given as  
\begin{equation}
  J_{\mu}^{(1b)} =  -n_{\mu} \int d\omega B_b (\omega)
\Bigl[\overline{\xi}_n W_n \delta  (\omega -\overline{n} \cdot
 \mathcal{P}^{\dagger}) \Bigr] \Bigl[W_n^{\dagger}
 i\FMSlash{D}_n^{\perp} W_n \Bigr]
\frac{1}{n\cdot \mathcal{P}} W_{\bar{n}}^{\dagger} \xi_{\bar{n}}
=-n_{\mu} j^{(1b)}, 
\end{equation}
where $B_b (\omega)$ is the Wilson coefficient, which is 1 at tree
level. The Feynman rules for $j^{(1b)}$ is given in Fig.~\ref{figjb},
and the Feynman diagrams for the radiative corrections are
given in Fig.~\ref{figjbr}. In Fig.~\ref{figjbr}, diagrams (a) to (e)
are the radiative corrections from the $n$-collinear loop
diagrams. Diagram (f) is from the $\overline{n}$-collinear loop
diagram, and diagrams (g), (h) are the contributions from the soft
loops. We employ the background gauge field method for the
triple-gluon vertex, and use the dimensional regularization with
$D=4-2\epsilon$. The external momenta $p^2$, $p^{\prime 2}$ and $q^2$
are kept to give infrared cutoff, and the poles in $1/\epsilon$ are of
the ultraviolet origin.

Considering the flow of momenta, $p^{\prime}-q$ is the total outgoing
collinear momentum in the $n^{\mu}$ direction, and $\overline{n}\cdot
(p^{\prime}-q) =Q$ is the large scale in DIS. We use the variables 
\begin{equation}
  \overline{n}\cdot (p^{\prime}-q)=Q, \ \overline{n}\cdot p^{\prime} =
  \omega^{\prime} = 
  Qv, \ \overline{n}\cdot q = \omega^{\prime} -Q = (1-v)Q, \ \omega =
  uQ, 
\end{equation}
and the allowed kinematic regions in DIS is $Q \geq \omega >0$. 
We extract the terms proportional to
$\gamma_{\perp\mu}$, and the divergent terms from each
category ($n$ collinear, $\overline{n}$ collinear, and soft) using the
dimensionless variables are given as
\begin{eqnarray} \label{cosum}
 iM_n &=& i(M_a +M_b + M_c +M_d + M_e)  \\
&=& \Gamma_{\mu} \Biggl[ \delta (u-v)  \Bigl\{ 2C_F
\Bigl(\frac{1}{\epsilon^2} +\frac{1}{\epsilon} \Bigr)
-\frac{N}{\epsilon} \Big(\ln \frac{-q^2}{\mu^2} +\ln u (1-u)
\Bigr)  +\frac{1}{N\epsilon} \ln \frac{-p^{\prime 2}}{\mu^2} \Bigr\}
\nonumber \\
&&+\frac{N}{\epsilon} \Bigl\{ \Bigl( \frac{1-u-v}{1-v} -u\Bigl) \theta
(u-v) +\Bigl( \frac{u(1-u-v)}{(1-u)v} -u\Bigr) \theta (v-u) \nonumber
\\ 
&&+ \frac{\theta (u-v)}{(u-v)_+} +\frac{u}{v} \frac{\theta
  (v-u)}{(v-u)_+} \Bigr\} \nonumber \\
&&+\frac{1}{N\epsilon} \Bigl\{ \Bigl( \frac{u(1-u-v)}{(1-u)(1-v)}
+x\Bigr) \theta (1-u-v) +\frac{(1-u)(1-v)}{v} \theta(u+v-1) \Bigr\}
\Biggr], \nonumber \\
iM_{\bar{n}} &=& iM_f =  \Gamma_{\mu} C_F \delta(u-v)
\Bigl(\frac{2}{\epsilon^2} -\frac{2}{\epsilon} \ln \frac{-p^2}{\mu^2}
+\frac{2}{\epsilon} \Bigr), \nonumber \\
iM_s &=& i(M_g+M_h) =  \Gamma_{\mu} \delta (u-v) \Biggl[
\frac{-2C_F}{\epsilon^2} 
+\frac{N}{\epsilon} \ln \frac{(-p^2)(-q^2)}{n\cdot p \overline{n}
  \cdot q \mu^2} -\frac{1}{N\epsilon} \ln \frac{(-p^2) (-p^{\prime
    2})}{n\cdot p \overline{n}\cdot p^{\prime} \mu^2} \Biggr],
\nonumber 
\end{eqnarray}
where $\Gamma_{\mu}$ is the common factor, given as
\begin{equation}
  \Gamma_{\mu} =\frac{\alpha_s}{4\pi} gT_a Q\frac{\gamma_{\perp
      \mu}}{n\cdot p}. 
\end{equation}
We can compute the above matrix elements using the zero-bin
subtraction. The infrared poles in $1/\epsilon_{\mathrm{IR}}$ cancel 
when we add the soft contributions and the zero-bin subtractions and
all the remaining poles turn into the ultraviolet poles. This
procedure is similar to the pullup mechanism in NRQCD
\cite{Hoang:2001rr}.    

The relation between the bare operator $j_B^{(1b)}$ and the
renormalized operator $j_R^{(1b)}$ is 
\begin{equation}
  j_R^{(1b)} (u) = \int_0^1 dv Z_B (u,v) j_B^{(1b)} (v), 
\end{equation}
where the operators are dimensionless operators expressed in terms of
$u$, $v$ instead of $\omega$ and $\omega^{\prime}$. The counterterm
$Z_B$ including the wave function renormalization is given by
\begin{eqnarray}
  Z_B (u,v) &=& \Biggl[ 1+\frac{\alpha_s C_F}{4\pi} \Bigl(
  \frac{2}{\epsilon^2} +\frac{3}{\epsilon} -\frac{2}{\epsilon} \ln
  \frac{uQ^2}{\mu^2} \Bigr) -\frac{\alpha_s}{4\pi} \frac{2}{N\epsilon}
  \ln (1-u) \Biggr] \delta (u-v) \nonumber \\
&+& \frac{\alpha_s}{4\pi} \frac{N}{\epsilon} \Biggl[ \Bigl(
\frac{1}{(u-v)_+} +1-u -\frac{u}{1-v} \Bigr) \theta (u-v) \nonumber \\
&&+\Bigl( \frac{u}{v} \frac{1}{(v-u)_+} +\frac{u}{v} -\frac{u}{1-u} -u
\Bigr) \theta (v-u) \Biggr]  \\
&+& \frac{\alpha_s}{4\pi} \frac{1}{N\epsilon} \Biggl[ u\Bigl(
\frac{1-u-v)}{(1-u)(1-v)} +1\Bigr) \theta (1-u-v)
+\frac{(1-u)(1-v)}{v} \theta (u+v-1) \Biggr]. \nonumber
\end{eqnarray}
Note that the mixture of the ultraviolet and infrared divergences such
as $[\ln (-q^2/\mu^2)]/\epsilon$ in Eq.~(\ref{cosum}) cancels when all
the contributions are summed. The renormalization group equation for
the current operator $j^{(1b)}$ is written as
\begin{equation}
  \mu\frac{d}{d\mu} j^{(1b)} (u) = -\int dv \gamma_B (u,v) j^{(1b)}
  (v), 
\end{equation}
where the anomalous dimension $\gamma_B (u,v)$ is given as
\begin{eqnarray} \label{gammab}
  \gamma_B (u,v) &=& Z_B^{-1} \Bigl( \mu \frac{\partial}{\partial\mu}
  +\beta \frac{\partial}{\partial g} \Bigr) Z_B (u,v)  \\
&=& -\Bigl[ \frac{\alpha_s C_F}{\pi} \Bigl(\frac{3}{2} -\ln
\frac{uQ^2}{\mu^2} \Bigr) -\frac{\alpha_s}{\pi} \frac{1}{N} \ln (1-u)
\Bigr] \delta (u-v) \nonumber \\
&-& \frac{\alpha_s}{\pi} \frac{N}{2} \Bigl[ \Bigl( \frac{1}{(u-v)_+}
+1-u-\frac{u}{1-v} \Bigr) \theta (u-v) \nonumber \\
&&+\Bigl( \frac{u}{v}
\frac{1}{(v-u)_+} +\frac{u}{v} -\frac{u}{1-u} -u \Bigr) \theta (v-u)
\Bigr] \nonumber \\
&-&\frac{\alpha_s}{\pi} \frac{1}{2N} \Bigl[ u\Bigl(
\frac{1-u-v}{(1-u)(1-v)} +1\Bigr) \theta (1-u-v) +\frac{(1-u)(1-v)}{v}
\theta (u+v-1)\Bigr]. \nonumber 
\end{eqnarray}

Those terms in Eq.~(\ref{gammab}) proportional to $\delta (u-v)$ come
from all the contributions, but the remaining terms proportional to
the theta functions originate from the $n$-collinear radiative
corrections. Compared to the renormalization of the subleading
heavy-collinear currents in Ref.~\cite{Hill:2004if}, the contributions
of the $n$-collinear radiative corrections are the same because the
contributing Feynman diagrams are the same. But the
soft and the $\overline{n}$ contributions should be different due to
the difference of the back-to-back collinear current and the
heavy-to-collinear current. Specifically the
contributions not proportional to $\delta (u-v)$ in our computation
and in Eq.~(\ref{gammab}) are the same. For $J_{\mu}^{(1a)}$, the
radiative corrections can be obtained in the 
same way as in the case of $J_{\mu}^{(1b)}$, and it satisfies
Eq.~(\ref{jab}).

\end{document}